\documentclass[11pt]{article}
\usepackage{jheppub} % for details on the use of the package, please see the JINST-author-manual
\usepackage[capitalize]{cleveref}
\usepackage{amsmath, bbold}
\usepackage{appendix}
\usepackage[dvipsnames]{xcolor}
\usepackage[normalem]{ulem}
\usepackage{graphicx, float}%subfigure, subcaption
\usepackage{tikz}
\usetikzlibrary{decorations.pathmorphing, positioning, calc, arrows.meta, math}
\usepackage{pgfplots}
\usepgfplotslibrary{fillbetween}
\pgfplotsset{compat=1.18, set layers}
\graphicspath{{./figures/}}

\usepackage[font=small,labelfont=sc]{caption}
\newcommand{\SU}[1]{\mathrm{SU}(#1)}

\newcommand{\U}[1]{\mathrm{U}(#1)}

\newcommand{\geff}{\rho}
\newcommand{\gem}{g_{m}^\textsc{em}}
\newcommand{\gemD}{g_\textsc{d}^\textsc{em}}
\newcommand{\gD}{g_\textsc{d}}
\newcommand{\Mvac}{\mathcal{M}_{\rm vac}} % YZ: a symbol I use for the vacuum manifold
\newcommand{\UIem}{\U{1}_{\mathrm{EM}}} % YZ: em \U{1}
\newcommand{\UIP}{\U{1}_{\mathrm{P}}} % YZ: primordial \U{1}
\definecolor{darkred}{rgb}{0.6,0,0}
\definecolor{darkpurple}{rgb}{0.5,0,0.5}

\newcommand{\TRH}{T_\text{max}}

\def\z2{$\mathbb{Z}_2$}
\def\z3{$\mathbb{Z}_3$}
\def\321{$SU(3)_c \times S\U{2}_L \times \U{1}_\mathrm{Y}$}

\def\555{\ensuremath{SU(5)^3}}

\def\24{\ensuremath{\mathbf{24}}}

\newcommand{\GeV}{~\mathrm{GeV}}
\newcommand{\TeV}{~\mathrm{TeV}}
\newcommand{\vlq}{v_{\rm NP}}

\newcommand{\commentout}[1]{} % YZ: a command I use for commenting out multiple lines of LaTeX code.

\author[a, b]{Marko Pesut,}
\author[c, b, d]{Davide Racco}
\author[e, d]{and Yunlong Zhao}

\affiliation[a]{PSI Center for Neutron and Muon Sciences,
5232 Villigen PSI, Switzerland}
\affiliation[b]{Physik-Institut, Universit\"at Z\"urich, Winterthurerstrasse 190, 8057 Z\"urich, Switzerland}
\affiliation[c]{Dipartimento di Matematica e Fisica, Università degli Studi Roma Tre and\\
INFN Sezione di Roma Tre, via della Vasca Navale 84, 00146 Rome, Italy}
\affiliation[d]{Institut f\"ur Theoretische Physik, ETH Z\"urich, Wolfgang-Pauli-Strasse 27, 8093 Z\"urich, Switzerland}
\affiliation[e]{Departement Physik, Universität Basel, Klingelbergstrasse 82, 4056 Basel, Switzerland}
\emailAdd{marko.pesut@psi.ch}
\emailAdd{davide.racco@uniroma3.it}
\emailAdd{yunlong.zhao@unibas.ch}

\preprint{ZU-TH 19/26}

\title{Model-Building Implications from Light Magnetic Monopoles: A Case Study of Flavour Deconstruction
}

\abstract{We highlight a generic connection between extensions of the Standard Model featuring low-scale semi-simple embeddings of $\rm{U}(1)_{\rm EM}$ and the phenomenology of light magnetic monopoles in the early Universe. Combining cosmological, astrophysical, and direct-search constraints, we show that magnetic monopoles with masses far below the scale of Grand Unified Theories are generically excluded and require low-scale inflation to dilute their abundance, followed by reheating below the monopole-production scale. These constraints are particularly relevant for flavour non-universal models, which provide a well-motivated framework to address the hierarchical structure of Yukawa couplings while allowing new dynamics close to the TeV scale compatible with experimental bounds. In many of these constructions, the sequential breaking of semi-simple gauge groups through intermediate stages containing a $\rm{U}(1)$ factor generically leads to the production of light magnetic monopoles. We show that the parameter region naturally predicted by these models implies low-scale inflation, with reheating temperature as low as $10^{3}\text{--}10^4 \,\mathrm{TeV}$. These results significantly reduce the otherwise large allowed inflationary window and establish a direct connection between flavour physics and the thermal history of the early Universe: future evidence for a low-scale semi-simple embedding of $\rm U(1)_{\rm EM}$, in particular through flavour non-universal interactions, would point towards low-scale reheating, while probes of primordial gravitational waves could directly constrain the scale at which low-scale semi-simple groups can appear.}

\begin{document}
\maketitle

\section{Introduction}

The origin of the flavour structure of the Standard Model (SM) is one of the most important theoretical puzzles motivating physics beyond the SM (BSM) \cite{Altmannshofer:2024hmr}. Even though the flavour puzzle does not by itself point to a definite energy scale for New Physics (NP), an indication in this direction may come from the Higgs hierarchy problem, which motivates at least one layer of BSM physics at scales not far above the electroweak (EW) scale, in order to stabilise the Higgs mass against radiative corrections from arbitrarily high energies. Given the stringent bounds from indirect searches on new sources of flavour and CP violation, any new physics appearing near the TeV scale must feature a highly non-generic flavour structure in order to remain phenomenologically viable~\cite{Isidori:2010kg}. In particular, flavour-universal new physics is typically constrained up to scales as high as $\mathcal{O}(10^{5})$~TeV.
Among the frameworks capable of reconciling stringent flavour bounds with new physics close to the EW scale, the paradigm of Minimal Flavour Violation (MFV)~\cite{DAmbrosio:2002vsn} suppresses new contributions to flavour-changing neutral currents (FCNCs) and separates the origin of flavour from the dynamics responsible for stabilising the Higgs sector. In this framework, the solution to the flavour puzzle is postponed to scales well above those motivated by the hierarchy problem. While MFV was particularly attractive in the pre-LHC era, the null results from direct searches encourage to relax the flavour assumptions of MFV~\cite{Davighi:2025icd,Isidori:2025iyu,Davighi:2023iks,Davighi:2025cqx,Allwicher:2023shc}. New physics coupling predominantly to third-family fermions and the Higgs can more easily evade collider constraints, allowing BSM states to remain closer to the EW scale~\cite{Allwicher:2023shc,Barbieri:2023qpf}. These constructions are motivated by the approximate $\U{2}^5$ global flavour symmetry observed in the Yukawa sector~\cite{Barbieri:2011ci,Barbieri:2012uh,Redi:2012uj,Isidori:2012ts,Fuentes-Martin:2019mun,Faroughy:2020ina}, which provides sufficient protection\footnote{The minimal symmetries compatible with TeV-scale new physics have recently been studied in~\cite{Greljo:2025mwj}.} against new sources of flavour violation, thereby allowing for new physics which does not worsen the Higgs hierarchy problem~\cite{Farina:2013mla}.

The paradigm of flavour non-universality, motivated by phenomenological constraints and the hierarchical structure of the Yukawa couplings, naturally points towards multiple layers of new physics associated with symmetry breaking at different energy scales. In this class of models dubbed Flavour Deconstruction~\cite{Li:1981nk,Arkani-Hamed:2001nha, Dvali:2000ha, Panico:2016ull,Bordone:2017bld, Allwicher:2020esa, Barbieri:2021wrc, Davighi:2022fer, Davighi:2022bqf, Davighi:2023evx, Fuentes-Martin:2022xnb, Barbieri:2023qpf, FernandezNavarro:2023hrf, Greljo:2018tuh, Fuentes-Martin:2020pww,FernandezNavarro:2024hnv,Craig:2011yk,Pesut:2025ook,Fuentes-Martin:2020bnh,FernandezNavarro:2022gst,Capdevila:2024gki,Fuentes-Martin:2024fpx, Greljo:2024ovt, Isidori:2025rci}, the flavour problem is addressed through a sequence of symmetry-breaking steps from initial flavour non-universal gauge groups to progressively more family-universal symmetries, ultimately recovering the SM gauge group\footnote{
An alternative possibility is to gauge horizontal flavour symmetries that commute with the SM gauge group, in the spirit of Froggatt--Nielsen models~\cite{Froggatt:1978nt} (see e.g.~\cite{DAgnolo:2012ulg,Linster:2018avp,Greljo:2023bix,Darme:2023nsy}). While such constructions can account for the Yukawa hierarchies, the extended gauge sector can be parametrically decoupled from the SM and is therefore subject to weaker phenomenological constraints. Moreover, the breaking of horizontal symmetries with no overlap with the SM gauge group does not generically lead to topological defects carrying electromagnetic charge, such as magnetic monopoles that are the main focus of this work. Nevertheless, as we discuss in Section~\ref{sec:implications}, our analysis still partially constrains the presence of monopoles in this class of models.}. Light fermion families acquire their Yukawa couplings at a high scale, where flavour non-universality is manifest, resulting in a parametric suppression relative to the third-generation Yukawa couplings, which are generated at lower scales. 
This intrinsically hierarchical construction separates the scales of new physics breaking $\U{2}^5$ and generating light-fermion masses (subject to strong flavour constraints) from the physics generating third-family masses and stabilising the Higgs sector, which can naturally lie near the TeV scale.
The phenomenology of flavour-deconstructed models at colliders, as well as in flavour and electroweak precision observables, has been studied extensively in e.g. Refs. ~\cite{Faroughy:2016osc,Gripaios:2014tna,Greljo:2018tuh,
Aebischer:2022oqe,Allwicher:2023aql,Cornella:2021sby,Davighi:2023xqn,Davighi:2023evx,Allwicher:2024ncl,Allwicher:2025bub,Fuentes-Martin:2019ign,Fuentes-Martin:2020bnh,Fuentes-Martin:2020hvc,Bauer:2015knc,Crivellin:2017zlb}. 

As a hint towards the UV gauge structure of the SM, the fractional pattern of hypercharges is explained only \emph{a posteriori} through anomaly cancellation and, together with the near unification of gauge couplings at high scales, naturally points towards semi-simple extensions of the SM gauge symmetry~\cite{Georgi:1974sy,Fritzsch:1974nn, Dimopoulos:1981yj, Dimopoulos:1981zb, Allanach:2021bfe}. 
Within flavour non-universal frameworks, such semi-simple embeddings can be realised at intermediates steps and separately for the flavour-deconstructed gauge groups acting on different fermion families. 
Ultimately, these flavour-deconstructed gauge symmetries may be unified into a single flavour-universal semi-simple gauge group \cite{Allanach:2021bfe,Davighi:2022dyq,Gosnay:2026dye}, as is commonly proposed in grand unified theories (GUTs).
\begin{figure}[t]\centering
\begin{tikzpicture}[declare function = {xGW=-3.; xlab=5.3; gap=0.5;
  yBBN=-2; yCol=2.7; yG3=4.1; yG12=6.; ycut=10.5; yr=12; 
  ymuHz=0; ymHz=3; yHz=6; ykHz=9;}]
\begin{axis}[
    width=9cm, height=9cm, clip=false,
    xmin=0, xmax=9.0,
    ymin=yBBN-0.5, ymax=yr+1,
    xtick=\empty,
    ytick={yBBN,yCol,yG3,yG12,yr},%Log10 GeV
    yticklabels={$10^{-2}$, $10^3$, $10^{3-4}$, $10^{5-6}$, $10^{15}$},
    ymajorgrids=true,
    grid style=transparent,
    axis x line=none,
    axis y line=left,
    ylabel={Energy scale [GeV]},
    axis line style={line width=1.2pt, -{Stealth} }] 
    %on layer=axis background %%force arrow to top layer
% Gradient regions (properly in background layer)
\begin{pgfonlayer}{axis background}
% Bottom region (balanced, visible, white fade at top)
\shade[bottom color=orange!20!white!80,
    top color=white]
(axis cs:0,yBBN) rectangle (axis cs:9.0,yr);
% Middle region (muted purple with white fade)
\shade[bottom color=NavyBlue!50!white!90,
    top color=white]
(axis cs:0,yCol) rectangle (axis cs:9.0,yG12);
\end{pgfonlayer}
% Centered labels
\node[Red, anchor=center] at (axis cs:xlab,yr-.55)
{Max.~$\TRH$ after inflation ($r$)};
\addplot[orange!85!black,  very thick, dotted]
coordinates {(0,yr) (9.0,yr)};
\node[magenta!70!black, anchor=center] at (axis cs:xlab,yG12)
{$G^{[12]} \to H^{[12]} \times \U{1}^{[12]}$};
\addplot[magenta!70!black, very thick, dotted] coordinates {(0,yG12) ({xlab-2-gap},yG12)};
\addplot[magenta!70!black, very thick, dotted] coordinates {({xlab+2+gap},yG12) (9.0,yG12)};
\node[purple!75!black, anchor=center] at (axis cs:xlab,yG3) {$G^{[3]} \to H^{[3]} \times \U{1}^{[3]}$};
\addplot[purple!75!black, very thick, dotted] coordinates {(0,yG3) ({xlab-1.8-gap},yG3)};
\addplot[purple!75!black, very thick, dotted] coordinates {({xlab+1.8+gap},yG3) (9.0,yG3)};
\node[teal!70!black, anchor=center] at (axis cs:xlab,yCol+.4){Lower bounds from colliders};
\addplot[teal!70!black,very thick, dotted] coordinates {(0,yCol) (9.0,yCol)};
\node[Red, anchor=center] at (axis cs:xlab,yBBN+.4){Min.~$\TRH$ after inflation (BBN)};
\addplot[orange!75!black, very thick, dotted] coordinates {(0,yBBN) (9.0,yBBN)};
% Points (clean triangular geometry)
\addplot[only marks, mark=*, mark size=3.4pt, red!85!black] coordinates {(1.58,yG12-.35)
    (1.25,yG12-.8)
    (1.72,yG12-1.05)};
\addplot[only marks, mark=*, mark size=3.4pt, green!60!black] coordinates {(1.38,yG3-.35)
    (1.05,yG3-.8)
    (1.52,yG3-1.05)};
% Arrows touching dashed lines
\draw[Red, thick, ->] (axis cs:2,yr) -- (axis cs:2,yr-.9);
\draw[Red, thick, ->] (axis cs:1.8,yBBN) -- (axis cs:1.8,yBBN+.9);
\draw[teal!70!black, thick, ->] (axis cs:2,yCol) -- (axis cs:2,yCol+.9);
% Axis break
\begin{pgfonlayer}{axis foreground}
% erase small part of axis (slightly thicker than axis)
\draw[line width=2.6pt, white] (axis cs:0, ycut+.08) -- (axis cs:0,ycut-.08);
% centered short slashes (axis passes exactly through midpoint)
\draw[line width=1.2pt] (axis cs:-0.2, ycut+0.2) -- (axis cs:0.2, ycut-0.);
\draw[line width=1.2pt] (axis cs:-0.2, ycut-0.) -- (axis cs:0.2, ycut-0.2);
% axis for GWs
\draw[very thick, -Stealth] (xGW,yBBN-0.5) -- (xGW,yr+1);
\draw[thin] (xGW-.15,ymuHz) -- (xGW,ymuHz);
\draw[thin] (xGW-.15,ymHz) -- (xGW,ymHz);
\draw[thin] (xGW-.15,yHz) -- (xGW,yHz);
\draw[thin] (xGW-.15,ykHz) -- (xGW,ykHz);
\node[left] at (axis cs:xGW-.1,ymuHz){$\mu$Hz};
\node[left] at (axis cs:xGW-.1,ymHz){mHz};
\node[left] at (axis cs:xGW-.1,yHz){Hz};
\node[left] at (axis cs:xGW-.1,ykHz){kHz};
\node[left, align=center] at (axis cs:xGW-.1,yr){Primordial\\ GWs};
\node[rotate=90,right] at (axis cs:xGW-2,ymHz-1){\textcolor{Mulberry}{\textbf{LISA}}};
\node[rotate=90,left] at (axis cs:xGW-2,ykHz){\textcolor{Mulberry}{\textbf{ET,CE}}};
\node[rotate=90,left] at (axis cs:xGW-2,ymuHz-.5){\textcolor{Mulberry}{\textbf{PTA}}};
\end{pgfonlayer}
\end{axis}
\end{tikzpicture}
\caption{Flavour non-universal models typically predict semi-simple gauge groups acting on the third generation ($G^{[3]}$) at scales around $10^{3}$--$10^4\,\GeV$, as well as additional factors acting on the first two families ($G^{[12]}$) at higher scales in order to reproduce the observed Yukawa hierarchies and suppress flavour violation in the light sector. Each symmetry-breaking step involving a semi-simple factor broken to a $\U{1}$ produces monopoles, which, as we show in this work, are phenomenologically excluded unless inflation dilutes their abundance and reheating occurs below the corresponding symmetry-breaking scale. In flavour-deconstruction models, and more generally whenever semi-simple embeddings of $\UIem$ are broken at low scales, constraints from light magnetic monopoles significantly reduce the otherwise large allowed inflationary window and establish a direct connection between BSM (flavour) physics and the thermal history of the early Universe. As an example, a future detection of primordial gravitational waves would strongly disfavour semi-simple gauge structures below the corresponding reheating scale.}
\label{fig:sketch}
\end{figure}
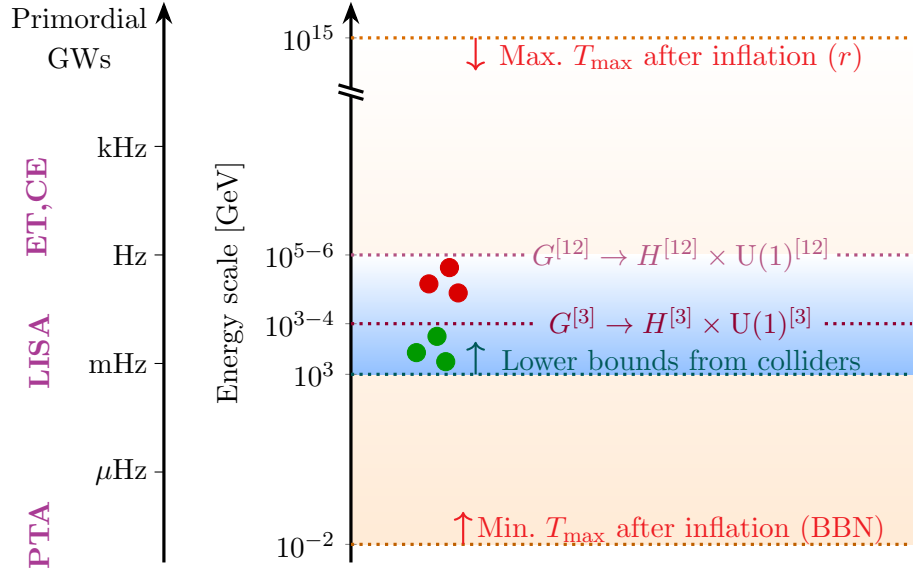

An important consequence of a cosmological phase transition in which a semi-simple gauge group is spontaneously broken to the SM (or to an intermediate gauge group containing a $\U{1}$ factor) is the production of monopoles, topologically stable field configurations \cite{tHooft:1974kcl,Polyakov:1974ek}, carrying magnetic charge under electromagnetism (EM). In the cosmological budget, they redshift as matter and can overclose the Universe \cite{Zeldovich:1978wj,Preskill:1979zi}. A GUT completion of the SM is subject to this threat, and the monopole problem was historically among the seminal motivations for primordial inflation \cite{Guth:1980zm,Linde:1981mu} well before many of the inflationary predictions were confirmed. 
An epoch of exponential dilution of all relic species, including monopoles, effectively erases them from our observable universe, as long as the Universe reheats below the associated symmetry-breaking scale. Other solutions to the monopole problem have been proposed (see e.g. Refs. \cite{Langacker:1980kd, Linde:1980tu, Dvali:1995cj, Dvali:1997sa, Bajc:1998rd, Stojkovic:2004hz, Perri:2025vtn}), at the cost of introducing further BSM structure and hypotheses. 
In this paper, we rely on primordial inflation as a highly motivated and economical solution to solve the monopole problem.

The potential production of monopoles has long been studied in the context of GUT theories~\cite{Zeldovich:1978wj,Preskill:1979zi, Dokos:1979vu, Jeannerot:2003qv, Murayama:2009nj, King:2021gmj}, where the symmetry-breaking scale at which monopoles are formed lies close to the maximal reheating temperature inferred from the current CMB bound on primordial tensor modes~\cite{BICEP:2021xfz,Tristram:2021tvh,BICEP2:2014owc}, corresponding to $T_{\rm max}\lesssim 10^{15}\,\GeV$ assuming an instantaneous reheating.
From this point of view, the requirement of diluting monopoles via an inflationary epoch adds little information on the physics responsible for inflation, and on the earliest cosmological history.

The main point of our work is that semi-simple embeddings of $\UIem$ broken to $\U{1}$ factors at scales $v_m\lesssim \mathcal{O}(10^{3})\TeV$, well below the GUT scale, lead to the production of light magnetic monopoles. We show that cosmological and, more importantly, astrophysical constraints on such monopoles are substantially stronger than the overclosure bound and therefore severely constrain these scenarios, unless inflation dilutes the monopole abundance and reheating occurs below the monopole-production scale.
In contrast to the conventional GUT monopole problem, this requirement removes a large fraction of the otherwise allowed reheating window and therefore imposes a qualitatively stronger constraint on the inflationary history of the Universe. These bounds are specifically relevant for BSM flavour models, in particular flavour non-universal gauge extensions, which, as summarised in \cref{fig:sketch}, typically feature low-scale breakings of semi-simple gauge groups\footnote{The necessity of diluting monopoles in flavour models was also noted in Refs. \cite{Greljo:2019xan,Cordova:2022qtz}.}. While our analysis will mostly focus on this class of extensions, we stress that our results apply more generally to any model featuring the low-scale breaking of semi-simple gauge embeddings of the $\UIem$ gauge symmetry.

The implications of these findings extend in both directions between particle physics and cosmology. Any future indication from direct or indirect searches for flavour non-universal dynamics, or more generally for low-scale semi-simple embeddings of $\UIem$, would point towards an inflationary history in which reheating occurred not far above collider scales. Such a low reheating scale would suggest either a less generic, potentially tuned, model for inflation or an inefficient reheating mechanism.
Conversely, future probes of primordial gravitational waves could provide direct indications on the highest scale at which semi-simple (flavoured) gauge groups may appear. In particular, the detection of primordial $B$-modes or of a stochastic gravitational-wave background%
\footnote{Related connections between flavour symmetry breaking, first-order phase transitions, cosmic defects and gravitational-wave signatures have recently been explored in Refs. \cite{Greljo:2019xan,Blasi:2024vew,Antusch:2025xrs,Fabri:2025fsc,Chrysostomou:2025vrg,Antusch:2026msw}.}
at frequencies above the mHz range, accessible to experiments such as LISA~\cite{LISA:2017pwj,LISACosmologyWorkingGroup:2022jok}, Einstein Telescope (ET)~\cite{ET:2025xjr,Punturo:2010zz,Hild:2010id}, Cosmic Explorer (CE)~\cite{Reitze:2019iox}, and atom-interferometer proposals such as MAGIS~\cite{Graham:2017pmn,MAGIS-100:2021etm,Graham:2012sy} and AION~\cite{Badurina:2019hst} as well as other intermediate-band proposals \cite{Crowder:2005nr,Kawamura:2011zz,AEDGE:2019nxb,Tino:2019tkb,MAGIS-100:2021etm}, would point towards reheating temperatures above the corresponding scale and therefore disfavour the presence of semi-simple gauge factors below it. Our study thus complements traditional phenomenological approaches to flavour-motivated BSM extensions by establishing a direct connection with the inflationary history of the Universe.

The paper is organised as follows. In \cref{sec:monopoles flavour}, we introduce topological monopoles and illustrate how they can be generically produced far below the GUT scale, with flavour-deconstructed models providing a particularly well-motivated example. Their cosmological evolution and the corresponding observational constraints are discussed in \cref{sec:monopoles pheno}, while the implications for flavour model building and early-Universe cosmology are presented in \cref{sec:implications}. Finally, we summarise our findings in \cref{sec:conclusions}.

\section{Monopoles in flavour extensions of the SM}
\label{sec:monopoles flavour}

In this section, we briefly review topological monopoles and argue that a broad class of flavour-motivated BSM completions can naturally accommodate them at scales not far above those probed by current collider experiments.

\subsection{Topological monopoles} 
\label{subsec:intro to monopoles}

We consider the spontaneous breaking of a gauge group $G \to H$, induced by a scalar field $\phi$ acquiring a vacuum expectation value (vev). 
The set of degenerate vevs defines the vacuum manifold $\Mvac \equiv G/H \cong \left\{\phi \mid V(\phi)=V_{\min}\right\}$, whose topology can be characterised by its homotopy groups $\pi_n(\Mvac)$. 
These groups consist of equivalence classes of maps from an $n$-dimensional sphere $S^n$ to $\Mvac$ that cannot be continuously deformed into one another. Topological monopoles, which are stable point-like solutions to the classical field equations, exist when the second homotopy group $\pi_2(\Mvac)$ is non-trivial, namely when $\pi_2(\Mvac)\neq I$, with $I$ denoting the trivial group (see also Ref. \cite{Shnir:2005xx} for an introduction and Refs. \cite{Mavromatos:2020gwk,Mitsou:2026zcf} for recent reviews). If such a symmetry breaking induces a phase transition in the early universe, cosmological production of monopoles occurs via the Kibble-Zurek mechanism \cite{Kibble:1976sj,Zurek:1985qw}.

In this paper, we focus on the symmetry breaking pattern 
\begin{equation}
\label{eq:monopolestep}
G \to K \times \UIP\,, 
\end{equation}
where $G$ and $K$ are both semi-simple gauge groups.
The subscript $\rm P$ on the $\U{1}$ factor denotes ``primordial'', since we generally take it as a precursor of the electromagnetic gauge group $\UIem$.
In this case, it can be shown that $\pi_2(\Mvac)=\pi_2(G/\left(K \times \UIP\right))\neq I$, and thus monopoles are always allowed\footnote{See Appendix~\ref{app:proof} for a proof. For this statement to hold, we also require $G$ to be compact, as is generally the case for gauge groups. Other mechanisms for producing topologically stable monopoles have been explored in Ref.~\cite{Lazarides:2024niy}.}.

Let us now turn to the simplest case of 't Hooft--Polyakov monopoles \cite{tHooft:1974kcl,Polyakov:1974ek}, where $G=\SU{2}$, $K=I$, and the symmetry is broken by the vev of an $\SU{2}$ scalar triplet. The mass of the unit winding monopoles with magnetic charge $g_{m}^{\text{'tH--P}}=\pm 4\pi/g$, where $g$ is the $\SU{2}$ gauge coupling, is given by
\begin{equation}\label{eq:monopole_mass}
    m = \frac{4\pi v_m}{g} \, f\!\left(\frac{\lambda}{g^2}\right)\,,
\end{equation}
where $v_m$ and $\lambda$ are the vev and quartic self-coupling of the scalar triplet. The function $f(x)$ lies in $[1,1.787)$ for $x \in [0,\infty)$ \cite{Bogomolny:1976ab}. Therefore, the monopole mass is nearly insensitive to $\lambda$ and determined only by $g$ and $v_m$.

In the more general case where the semi-simple factor $K$ is nontrivial, we expect this formula to remain approximately valid, since as it suggests, the monopole mass is dominated by the scalar gradient energy within its core of size $R_m\sim (g v_m)^{-1}$ (see e.g. Ref. \cite{Preskill:1984gd} for detailed discussions). This classical description of monopoles remains applicable as long as the core size exceeds the monopole Compton wavelength $R_m\gtrsim \lambda_m \sim m^{-1}$, which breaks down when $g\gtrsim \sqrt{4\pi}$.

When the gauge group $K\times \UIP$ is subsequently broken through a symmetry breaking chain down to the SM and, after electroweak symmetry breaking (EWSB), to $\SU{3}_\mathrm{c} \times \UIem$, the primordial monopoles may split, cluster, or even annihilate during the intermediate symmetry breakings. However, as long as $\UIP$ has a nontrivial overlap with $\UIem$, the true vacuum manifold satisfies $\pi_2\!\left(G/(\SU{3}_\mathrm{c} \times \UIem)\right) \neq I$, implying that electromagnetic monopoles are topologically allowed and that a nonzero fraction of the primordial monopoles is expected to survive and eventually become EM magnetic monopoles. The Dirac quantization condition then applies to their $\UIem$ magnetic charge $g^{\rm EM}_m$, yielding
\begin{equation}\label{eq:g_m^EM}
    g^{\rm EM}_m=\frac{2\pi k_{\rm EM}}{e},\quad k_{\rm EM} \in \mathbb{Z} \backslash\{0\} \,,
\end{equation}
where $e$ is the EM gauge coupling.

\subsection{Examples of monopole-producing flavour-deconstruction models}
\label{subsec:FD model examples}

The goal of this work is to connect monopole phenomenology to well-motivated BSM scenarios, such as flavour-motivated gauge extensions. These constructions can naturally generate monopoles at symmetry-breaking steps of the type shown in \cref{eq:monopolestep}, which arise generically in flavour models based on semi-simple UV completions of extended, and possibly flavour non-universal, gauge symmetries. We emphasize that the phenomenology of light monopoles and the connection to low-scale inflation discussed in this paper applies generically to any model featuring a low-scale semi-simple embedding of $\UIem$, and is not specific to the framework of flavour deconstruction. Nevertheless, this framework provides a motivated example of BSM theories
where semi-simple unification can occur, in a possibly family-dependent manner and hence
multiple times along the full symmetry-breaking chain, at scales significantly lower than the
GUT scale, while remaining compatible with current experimental bounds.

In this class of models, the hierarchies in the Yukawa couplings are generated through a multi-scale structure, where flavour dynamics associated with high-scale symmetry breaking produce suppressed Yukawa entries for the light families, while the larger third-family Yukawa couplings originate from dynamics not far above the TeV scale. Consequently, the multi-scale structure of these models is intrinsically tied to the solution of the flavour puzzle and cannot be arbitrarily adjusted.
 
Flavour and collider constraints generically push new physics breaking the approximate $\U{2}$ flavour symmetries of the SM to scales of $\mathcal{O}(10^2\text{–}10^5)$ TeV, still well below the GUT scale, while TeV-scale $\U{2}$-preserving dynamics remain compatible with direct and indirect searches as long as they dominantly couple to third-family fields \cite{Allwicher:2023shc}. 
On top of addressing the flavour puzzle of the SM through new flavour-deconstructed gauge symmetries, these models have the appealing feature of providing a flavour structure compatible with new physics at scales close to the EW scale, and hence can be elegantly combined with a natural solution to the Higgs hierarchy problem \cite{Covone:2024elw,Davighi:2025cqx,Glioti:2024hye}. An exhaustive review of the different features of flavour-deconstructed
extensions is outside the scope of this work, and we refer the reader to
e.g. Ref. \cite{Davighi:2023iks} and references therein for further details.

In the remainder of this section, we provide three representative examples of flavour non-universal models featuring symmetry-breaking steps that produce monopoles. 
Ref.~\cite{Davighi:2022fer} explores the idea of gauge-flavour unification \cite{Davighi:2022vpl} into
\begin{equation}
\SU{4} \times \mathrm{Sp}(6)_\mathrm{L} \times \mathrm{Sp}(6)_\mathrm{R} \,,
\end{equation}
which is then broken through a cascade of SSB steps. In particular, one of the breaking steps featured in the model, namely
\begin{equation}
\SU{4} \times \mathrm{Sp}(6)_\mathrm{R} \rightarrow \SU{3} \times \mathrm{Sp}(4)_\mathrm{R}^{[12]} \times \U{1}_\mathrm{R} \,,
\end{equation}
where $\mathrm{U}(1)_\mathrm{R}$ acts as hypercharge for the third family and for the left-handed fermions, is precisely like \cref{eq:monopolestep}. More precisely, since the surviving hypercharge generator contains the
$\U{1}_\mathrm{R}$ generator and because $Q_{\rm EM}=T_{\mathrm L}^3+Y$, this breaking step produces monopoles that will acquire a non-vanishing EM
magnetic charge after EWSB, as indicated also by the authors in Ref. \cite{Davighi:2022fer}. Moreover, this SSB step produces massive flavour-universal vector leptoquarks (LQs) from the breaking of $\SU{4}$, and should  therefore occur at scales $\mathcal{O}(10^2)\,\mathrm{TeV} \lesssim \Lambda \ll \Lambda_{\rm GUT}$, where the lower bound on $\Lambda$ is set by flavour-changing processes\footnote{In models based on the Pati--Salam gauge group, in contrast to GUT models based on $\SU{5}$, vector gauge bosons do not mediate proton decay.} 
involving light families which receive additional contributions from LQ exchange, such as $K_L \to \mu e$. 

Another example is provided by the model studied in Ref. \cite{Davighi:2022bqf}, which also features the idea of gauge–flavour unification, although in this case it is applied only to the light families. The UV gauge group consists of 
\begin{equation}
\left[\SU{4} \times \mathrm{Sp}(4)_\mathrm{L} \times \mathrm{Sp}(4)_\mathrm{R}\right]^{[12]} \times\left[\SU{4} \times \SU{2}_{\mathrm{L}} \times \SU{2}_{\mathrm{R}}\right]^{[3]}\,,
\end{equation}
which is spontaneously broken as
\begin{equation}
\left[\SU{4} \times \mathrm{Sp}(4)_\mathrm{R} \right]^{[12]}  \xrightarrow[]{} \SU{3}^{[12]} \times \SU{2}_{\mathrm{R}}^{[1]} \times \U{1}_\mathrm{R}^{\prime \prime},
\end{equation}
This breaking step is of the form given in \cref{eq:monopolestep} and produces monopoles at a scale $\Lambda \approx \mathcal{O}(10^3)$ TeV $\ll \Lambda_{\rm GUT}$, which is again constrained by flavour-changing processes involving the first two generations of quarks \cite{Isidori:2010kg}.

Finally, Ref.~\cite{Bordone:2017bld} considers a fully deconstructed Pati-Salam \cite{Pati:1974yy} UV extension
\begin{equation}
\mathrm{PS}^3 \equiv \mathrm{PS}^{[1]} \times \mathrm{PS}^{[2]} \times \mathrm{PS}^{[3]} \,,
\end{equation}
where each PS factor, which consists of $\SU{4}\times \SU{2}_\mathrm{L}\times \SU{2}_\mathrm{R}$, acts on a single fermion family. The first symmetry breaking step occurs at $\Lambda \gtrsim  \mathcal{O}(10^3)$ TeV and involves 
\begin{equation}
    \text{PS}^{[1]} \to \text{SM}^{[1]}\,,
\end{equation}
where $\text{SM}^{[1]}$ denotes SM gauge symmetries acting only on the first family of fermions\footnote{Although this scale is not directly fixed by the generation of the Yukawa hierarchies, it controls the decoupling of the first-family Pati--Salam gauge bosons and is bounded from below by flavour observables involving light families. Pushing it parametrically above the other deconstruction scales would amount to an additional hierarchy in the breaking chain, requiring either a model-building assumption or a protective mechanism.}. This step also falls within the class described in \cref{eq:monopolestep}, since the SM group contains $\U{1}_\mathrm{Y}$.

These three examples should be regarded as illustrative cases of the central message of this work: BSM extensions featuring the breaking of semi-simple embeddings of $\UIem$ at scales well below the GUT scale unavoidably lead to the production of light magnetic monopoles. In particular, models with enlarged, and possibly flavour non-universal, gauge symmetries aimed at addressing the flavour puzzle typically satisfy these requirements, with the corresponding symmetry-breaking scales tied to the observed Yukawa hierarchies and constrained by the absence of new sources of flavour violation. When these symmetries are spontaneously broken to explicit $\U{1}$ factors, usually with a non-trivial overlap with $\U{1}_{\rm EM}$, the formation of magnetic monopoles is generically unavoidable.

\subsection{Constraints on flavour-deconstructed models}
In \cref{subsec:FD model examples}, we identified the conditions under which a SSB step produces monopoles and pointed out that such symmetry breakings are particularly common in flavour models, especially in BSM extensions featuring flavour non-universal gauge symmetries. 
The theoretical interest in these models has mostly focused on the lowest-energy layer of new physics, that we denote by $\vlq$ (typically around a few TeV), which can provide detectable signals in current and upcoming experiments. 
%In particular, it typically features 
TeV-scale massive gauge bosons with flavour non-universal interactions %, which 
can leave observable signatures in electroweak precision tests, direct searches, and flavour observables. We focus on two generic bounds capturing the main features of a broad class of such models.

\begin{enumerate}

\item \textbf{Naturalness:} As a representative criterion inspired by finite naturalness considerations \cite{Farina:2013mla}, one may require that the Higgs mass correction induced by the new massive gauge bosons,
\begin{equation}
\label{eq:constraint naturalness}
\delta m_H^2 \sim \frac{g^2 M^2}{16\pi^2} \lesssim (1\mathrm{\,TeV})^2 \,,
\end{equation}
does not induce more than a percent-level tuning in the Higgs sector. For definiteness, we take $M \simeq \tfrac{1}{2}\, g\, \vlq$, up to $\mathcal{O}(1)$ prefactors arising from the running of $g$ and group-theoretical factors associated with the spontaneous symmetry breaking. This relation should be understood merely as a proxy for the parametric scaling between the gauge-boson mass and the symmetry-breaking scale, up to a numerical coefficient determined by vev ratios and combinations of gauge couplings. Since our discussion is intended to remain agnostic about the specific symmetry-breaking step, such model-dependent details are not relevant for our purposes.

\item \textbf{Bounds from direct searches:} In flavour-deconstructed models, the first layer of new physics typically consists of massive gauge bosons with enhanced couplings to the third family and suppressed, approximately $\U{2}$-preserving, couplings to the light families, which naturally avoids many of the most stringent flavour constraints.
LHC direct searches for heavy resonances coupled to $t,b,\tau$ set lower limits in the ballpark of $1.5\text{--}2$ TeV (see e.g. Refs. \cite{ATLAS:2023vxj,CMS:2023qdw}), which can be naturally accommodated in flavour-deconstructed models featuring states dominantly coupled to third-family fields. Later in \cref{fig:summary plot}, we will choose as a reference 
\begin{equation}
\label{eq:constraint collider}
M=\frac 12\, g\, \vlq > 2\TeV \,.
\end{equation}
\end{enumerate}
The two conditions \cref{eq:constraint naturalness,eq:constraint collider} suggest an expected upper and lower limit on the theoretically motivated range for $\vlq$ (the former being a hint, and the latter a robust constraint). 

We stress that, while the signatures closest to experimental reach are typically associated with the last SSB step leading to the SM, $\vlq$ does not necessarily coincide with a monopole-producing scale. However, the lowest layer of new physics is generically embedded into a larger semi-simple gauge group with coupling $g$, whose breaking at a scale $v_m$ produces monopoles. The scale $v_m$ can naturally lie a few orders of magnitude above the TeV scale, while still remaining far below the conventional GUT scale (see \cref{fig:sketch}).

In order to translate collider limits on flavoured gauge bosons into the $(g,v_m)$ plane, an explicit assumption about the hierarchy between $\vlq$ and $v_m$ is required. We argue that, in a broad class of models including flavour non-universal gauge extensions, the scale $v_m$ at which the SM is embedded into an underlying semi-simple gauge factor is naturally expected to lie not far%
\footnote{This expectation is generally realised in flavour non-universal models, but in some cases the physics addressing the hierarchy problem (e.g.~compositeness) might lie at an intermediate scale $v_{\rm HP}$ such that $v_{\mathrm{NP}} \sim \mathcal{O}(\mathrm{TeV})\lesssim v_{\mathrm{HP}} \ll v_m$. 
In such a scenario, naturalness does not require $v_m$ to be close to $v_{\rm NP}$ or $v_{\rm HP}$, and monopoles associated with the high-scale breaking may lie parametrically %above the scales directly connected to the Higgs naturalness problem and to the lowest layer of NP
above $v_\mathrm{NP}$, whose phenomenology is relevant for current and future experiments \cite{Davighi:2023iks,Davighi:2025cqx}.}
above $\vlq$, while the corresponding gauge coupling $g$ remains comparable to the gauge couplings at $\vlq$ and at the electroweak scale. This makes it possible to directly connect the flavour and collider constraints discussed in this Section with the cosmological and astrophysical constraints on monopoles, produced at $v_m$, discussed in the next Section.

%%%%%%%%%%%%%%%%%%%%%%%%%%%%%%%%
\section{Cosmological and astrophysical constraints on magnetic monopoles}
\label{sec:monopoles pheno}

The aim of this section is to briefly review the cosmological production of monopoles and derive limits on their parameter space from cosmological and astrophysical constraints in the absence of inflationary dilution. We discuss monopole production in \cref{sec:monopole production} and the main observational bounds in \cref{sec:monopole bounds}. The different monopole velocity regimes relevant for these constraints are discussed in more detail in Appendix~\ref{app:monopole details}.

%%%%%%%%%%%%%%%%%

\subsection{Cosmological production of monopoles}
\label{sec:monopole production}

The initial abundance of monopoles $n_i$ is rather difficult to estimate precisely in full generality. 
We focus here on the case of a second-order or weakly first-order phase transition, which is the natural expectation for our parameter space of interest, namely $g$, $\lambda\sim \mathcal O(0.1)$, as shown in e.g. Ref. \cite{Brummer:2025inh}.
In this case, monopoles form at an initial temperature $T_i$ close to the critical temperature $T_c \sim v_m$, at which a semi-simple group is spontaneously broken to a subgroup containing $\U1_{\rm P}$. The initial monopole yield $r_i$ is estimated to be\footnote{The monopole number density at formation $n_{i} \sim \xi^{-3}$ depends on the critical exponent $\nu$ characterising the correlation length of the Higgs field around the critical temperature, $\xi \propto [(T-T_c)/T_c]^{-\nu}$. 
The Landau-Ginzburg theory features $\nu=\tfrac 12$, and we assume this value for simplicity, following \cite{Murayama:2009nj}. Various QFT examples and condensed-matter systems provide examples of $\nu\approx \tfrac 23$ (see Fig.~1 and Refs.~in \cite{Murayama:2009nj}).}
\begin{equation}
r_{i} \equiv \frac{n_i}{T_c^3}
  \simeq \frac{\lambda T_{c}}{2 CM_{\mathrm{Pl}}} \,,
\label{eq:n2opt}
\end{equation}
where $\lambda$ is the scalar quartic self-coupling. We assume a radiation-dominated Universe with a Hubble rate $H=T^2/CM_{\rm Pl}$, with $M_{\rm Pl}$ the unreduced Planck mass, $C=\sqrt{45/4\pi^3g_*(T_i)}$ and $g_*$ denotes the number of relativistic degrees of freedom, evaluated here at the temperature $T=T_i$. 

After production, at temperatures $T<T_i$, the comoving monopole density can only be reduced through monopole-antimonopole pair annihilation, due to the topological stability of monopoles. More precisely, monopole-antimonopole pairs diffuse through the plasma and annihilate via the long-range magnetic force associated with the unbroken $\U{1}_{\rm P}$, until their mean free path exceeds the capture distance set by the magnetic Coulomb potential, at which point the annihilation process effectively freezes out. Assuming that $g_*$ remains constant during annihilation, the freeze-out monopole yield is estimated to be \cite{Preskill:1984gd} 
\begin{equation}
r_f \equiv \frac{n_f}{T_f^3}
  \simeq \frac{1}{\geff g_m^2}\left(\frac{4 \pi}{g_m^2}\right)^2 \frac{m}{C M_{\rm Pl}}\,,
\label{eq:FOyield}
\end{equation}
where $m$ is the monopole mass, $T_f= 16\pi^2m/(\rho^2 g_m^4)$ is the freeze-out temperature, $g_m$ is the magnetic charge of the monopole under $\UIP$, $\geff \simeq \frac{2 \pi}{9} \sum_i b_i\left(\frac{g_m q_i}{4 \pi}\right)^2 \ln \Lambda$ counts the number of particles carrying $\U{1}_{\rm P}$ charge $q_i$, $b_i=1 (1/2)$ corresponds to bosons (fermions), and $\ln \Lambda$ is the Coulomb logarithm with $\Lambda \sim g_m^4 / g_*$.

%%%%%%%%%%%%%%%%%%%%%%%%%%%%%%%%
\subsection{Observational constraints on monopoles}
\label{sec:monopole bounds}

In this section, we review the main cosmological and astrophysical constraints on the abundance of monopoles. 
We do not consider the bounds relying on the Callan-Rubakov effect~\cite{Rubakov:1981rg,Callan:1982au}, which suggests an annihilation cross-section between monopoles and nucleons of the order of the strong cross section. 
This effect would lead to bounds on the monopole flux (see the [\href{https://pdglive.lbl.gov//DataBlock.action?node=S028FA}{\textsc{pdg}}]) that are 1-2 orders of magnitude more severe \cite{Dimopoulos:1982cz,Kolb:1982si, Freese:1983hz} than the strongest bound we consider. 
Given that these bounds would not affect our conclusions, and partly because of the uncertainty possibly associated to the cross section of the effect (see e.g. Ref. \cite{Csaki:2021ozp} for recent developments), we choose not to consider those bounds in the following. 

\begin{figure}[t]\centering
\includegraphics[width=0.65\textwidth]{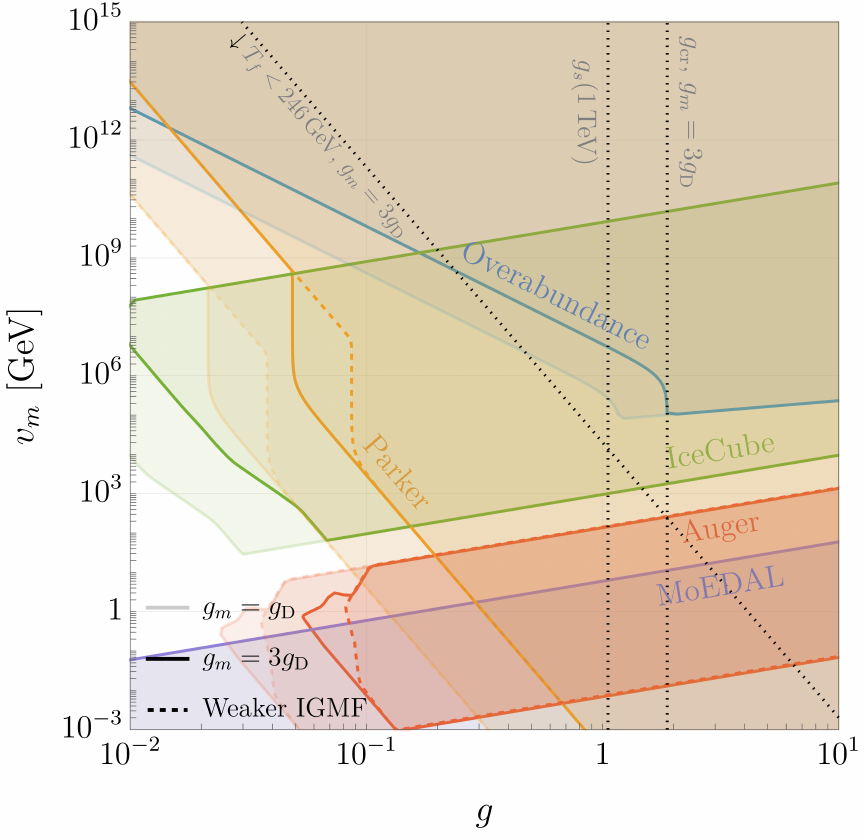}
\caption{Cosmological and astrophysical bounds on monopoles in the $(g,v_m)$ plane, assuming primordial magnetic charges $g_m=\gD=2\pi/g$ (light solid lines) and $g_m=3\gD$ (dark solid lines). We assume a one-to-one conversion between primordial and electromagnetic monopoles, with the latter carrying the minimal Dirac charge $\gem=\gemD$. Larger values of $\gem$ would generally strengthen the astrophysical bounds. As benchmarks for the IGMF compatible with current constraints~\cite{Durrer:2013pga}, following Ref.~\cite{Perri:2025qpg}, we take $B_{\rm I}=10^{-9}\,\mathrm{G}$ with $l_{\rm I}\gtrsim H_0^{-1}$, while dashed lines show the mild variation obtained for $B_{\rm I}=10^{-10}\,\mathrm{G}$ and $l_{\rm I}=1\,\mathrm{Mpc}$. Monopole velocities are estimated following Ref.~\cite{Perri:2025qpg} (see also Appendix~\ref{app:monopole details}). The critical coupling $g_{\rm cr}$ corresponds to $T_f\sim v_m$, such that to the right of $g_{\rm cr}$ the monopole abundance freezes out at formation and is set by its initial value, estimated assuming a second-order phase transition. We also set $g_*=200$ and neglect its mild temperature dependence. The regime in which pair annihilation continues after EWSB, $T_f \lesssim 246\,\GeV$, is indicated by a dotted line. We emphasize that these bounds apply generically to any model featuring the low-scale breaking of a semi-simple gauge group containing $\UIem$, and are not specific to flavour deconstruction.}
\label{fig:monopole astro}
\end{figure}

In the remainder of this section, we provide analytical expressions assuming that the magnetic charge of monopoles is large enough, so that their abundance is reduced by annihilations from $r_i$ to $r_f$. In \cref{fig:monopole astro}, we use the full expression, accounting for the possibility that the monopole abundance is frozen at formation.

\paragraph{Monopole overabundance}
The existence of a relic monopole abundance that is not diluted by inflation is a well-known problem in GUTs~\cite{tHooft:1974kcl,Polyakov:1974ek,Preskill:1979zi,Shnir:2005xx}: if they dominate the energy density before matter-radiation equality ($T_{\rm eq} \approx 1\,\mathrm{eV}$), they overclose the Universe. The requirement that the Universe is radiation-dominated at a temperature $T<T_f$ gives
\begin{equation}
r_f \lesssim \frac{\pi^2}{30} \frac{g_{*}(T_f)T}{m}.
\end{equation}

\paragraph{Parker bound}
Monopoles charged under a $\U{1}_{\rm P}$ factor with a non-trivial overlap with $\U{1}_{\rm EM}$ become electromagnetic magnetic monopoles, and they may deplete the galactic magnetic field (GMF) upon being accelerated by the intergalactic magnetic field (IGMF) permeating the Universe.
The existence of the GMF in the Milky Way (MW) implies an upper bound on the monopole flux, known as the Parker bound~\cite{Parker:1970xv,Turner:1982ag,Kobayashi:2023ryr}.

Depending on their mass, monopoles can either pass through or cluster with the galaxy. 
The latter occurs only for very heavy monopoles with $m\gtrsim 10^{18}\,\GeV$, well above the range relevant for our analysis, and this possibility is therefore neglected in the following.
For monopoles passing through the MW, their flux can be written as
\begin{equation}
F=\frac{n_{\cos } \beta_{\mathrm{MW}}}{4 \pi}\,,
\label{eq:funclustered}
\end{equation}
where $n_{\cos}=r_f \frac{g_{*,s}\left(T_0\right)}{g_{*,s}(T_f)} T_0^3$ is the present-day average number density  of monopoles in the Universe, and $\beta_\mathrm{MW}$ is the typical monopole velocity with respect to the MW. We refer to \cite{Kobayashi:2023ryr} for a derivation of the Parker bound, and we quote here their result on the upper limit on $F$, which is valid for $m\lesssim \mathcal O(10^{18})\GeV$:
\begin{equation}
\label{eq:flux parker}
F\lesssim 10^{-16} \mathrm{~cm}^{-2} \mathrm{~s}^{-1} \mathrm{sr}^{-1} \cdot 
  \left(\frac{\gemD}{\gem}
  \right)\left(\frac{B_{\mathrm{G}}}{10^{-6} \,\mathrm{G}}\right)\left(\frac{R}{l_{\mathrm{G}}}\right)^{1 / 2}\left(\frac{\tau_{\mathrm{gen}}}{10^8 \,\mathrm{yr}}\right)^{-1}\,,
\end{equation}
where $\gemD = 2\pi/e$ is the Dirac magnetic charge associated with EM, $B_{\rm G}$ denotes the strength of the GMF, $R$ the size of the Galaxy, $l_{\rm G}$ its coherence length, and $\tau_{\rm gen}$ the characteristic timescale over which the field is regenerated.

\paragraph{Direct searches for monopoles}
The most recent searches for magnetic monopoles range from searches for relativistic monopoles in cosmic rays (e.g.~IceCube and the Pierre Auger Observatory), to collider searches for monopole production (MoEDAL at the LHC), monopole-induced catalysed proton decay~\cite{Rubakov:1981rg,Callan:1982au}, and the accumulation of monopoles in bulk matter~\cite{Patrizii:2015uea}. The current lack of signals imposes upper limits on the monopole flux at the detector. We briefly review these bounds here.

\begin{itemize}
\item \emph{IceCube}: Located in the ice of the South Pole, it detects high-energy cosmic neutrinos via their conversion into charged leptons and the emission of Cherenkov radiation. The velocity threshold for Cherenkov radiation in ice is $\beta_{\mathrm{C}} = 1/\mathfrak{n}_{\mathrm r} = 0.76$, where $\mathfrak{n}_{\mathrm r}$ denotes the refractive index of ice.
IceCube is particularly sensitive to relativistic monopoles, as they emit a larger number of Cherenkov photons than those from a charged lepton by a factor $(\gemD\,\mathfrak{n}_{\mathrm r}/e)^2 \approx 8.2\times10^3$ for a given velocity~\cite{PhysRev.138.B248}.
Assuming an isotropic monopole flux, the most recent IceCube analysis \cite{IceCube:2021eye} sets an upper bound
\begin{equation}
F \lesssim 10^{-19}\text{--}10^{-17}\,\mathrm{cm^{-2}\,sr^{-1}\,s^{-1}}
\end{equation}
for velocities in the range $0.76 \lesssim \beta \lesssim 0.995$. This bound also applies to larger $\beta$, for which the Cherenkov emission is even stronger~\cite{IceCube:2016Monopoles}.

\item \emph{Pierre Auger Observatory}: 
Designed to detect cosmic rays via secondary electromagnetic showers and fluorescence light emitted by atmospheric nitrogen as they traverse the atmosphere, the experiment is also sensitive to ultra-relativistic magnetic monopoles. Auger data \cite{PierreAuger:2016imq} set an upper bound
\begin{equation}
F \lesssim10^{-21}\text{--}10^{-17}\,\mathrm{cm^{-2}\,sr^{-1}\,s^{-1}}
\end{equation}
on the (isotropic) monopole flux on Earth, for Lorentz factors $\gamma \gtrsim 10^{8}$ upon entering the atmosphere. 

\item \emph{MoEDAL}: The Monopole and Exotics Detector at the LHC (MoEDAL) \cite{MoEDAL:2014ttp} aims to detect monopoles produced at the LHC via the magnetic Schwinger mechanism. Originally proposed for the pair production of electric charges in a strong electric field~\cite{Schwinger:1951nm}, by EM duality, the magnetic Schwinger mechanism leads to monopole-antimonopole pair production in a strong magnetic field~\cite{Affleck:1981ag,Ho:2021uem,Gould:2019myj}. The magnetic field produced in heavy-ion collisions at the LHC can reach peak values $B \approx 10^{16}\,\mathrm{T}$ over time scales of $(73\,\mathrm{GeV})^{-1}$, providing an ideal environment for monopole production. The absence of a signal in MoEDAL translates into a lower bound on the monopole mass~\cite{MoEDAL:2021vix}
\begin{equation}
    m \gtrsim 75\,\mathrm{GeV} \,.
\end{equation}
\end{itemize}

In order to relate these bounds on monopole fluxes (which apply to different velocity regimes) to the monopole abundance%
\footnote{The number density of monopoles crossing the Galaxy can be taken to be approximately equal to $n_{\rm cos}$, since significantly altering the monopole flux on Galactic scales would require unrealistically large electric fields coherent over Galactic distances.}
and then to our parameter space of interest $(g, v_m)$ in \cref{fig:monopole astro}, one needs to estimate the typical velocity of monopoles as a function of their mass. We follow the estimate performed in Ref. \cite{Perri:2025qpg}, which we review in Appendix~\ref{app:monopole details}. 

For definiteness, following \cref{eq:monopole_mass}, we take the monopole mass to be $m= 4\pi v_m/g$ and use $g_m=\gD$ and $g_m=3\gD$ as benchmark values for the primordial magnetic charge, where $\gD\equiv 2\pi/g$ is the Dirac magnetic charge associated with $\U1_{\mathrm{P}}$. In doing so, we identify the gauge coupling of the group $G$, which sets the monopole mass in \cref{eq:monopole_mass}, with the gauge coupling of the residual $\U{1}_{\rm P}$ responsible for monopole annihilation, neglecting possible $\mathcal{O}(1)$ group-theoretical factors relating the two. 
Furthermore, the primordial-electromagnetic conversion of monopoles is model-dependent and thus difficult to quantify (see e.g. Ref. \cite{Lazarides:2023iim} for some examples). For definiteness, we assume a one-to-one conversion between primordial and electromagnetic monopoles, with the latter carrying a single Dirac charge, $\gem=\gemD$, which is the minimal allowed value of $\gem$. Larger values of $\gem$ would generally strengthen the astrophysical bounds because of the enhanced interactions between monopoles and magnetic fields. On the other hand, while increasing $g_m$ slightly weakens the bounds, as illustrated in \cref{fig:monopole astro}, monopole production is expected to be dominated by the minimally charged topological sector, which is energetically favoured. Monopoles carrying larger magnetic charge are therefore expected to be strongly suppressed. Finally, assuming that the particle content charged under $\U1_{\rm P}$ in the cosmic plasma during monopole annihilation is comparable to the EM sector of the SM, we set $\geff \simeq g_m^2/20$. We also take $g_*=200$ to be constant during monopole pair annihilation. This approximation necessarily breaks down when $T_f \lesssim 246\,\GeV$. However, a decrease in $g_*$ during annihilation would also reduce the monopole number density, thereby lowering the annihilation rate and strengthening the bounds. 

Under these assumptions, over a wide region of the parameter space shown in \cref{fig:monopole astro}, the monopole velocity on Earth is predominantly set by acceleration in the GMF. The only exception is a small region around $v_m\lesssim 1\,\GeV$ and $g \lesssim0.1$, where the IGMF dominates, as can be appreciated in \cref{fig_velregime} following the discussion in Appendix \ref{app:monopole details}.

The main message of \cref{fig:monopole astro} is that direct cosmic-ray and LHC searches, together with the Parker bound, essentially exclude light magnetic monopoles unless their abundance is diluted during the cosmological evolution. A possible exception exists only for monopoles originating from gauge groups with couplings $g \lesssim \text{few}\cdot 10^{-2}$ (white region in \cref{fig:monopole astro}), significantly smaller than the SM gauge couplings and well below the values typically expected in perturbative semi-simple extensions of the SM, where matching conditions generically imply gauge couplings of comparable size.

%%%%%%%%%%%%%%%%%%%%%%%%%%%%%%%%

\section{Impact of monopole constraints on flavour model building}
\label{sec:implications}
We can now discuss the implications of monopole production in extensions of the SM gauge group featuring the spontaneous breaking of semi-simple gauge factors at scales not far above the TeV scale, as motivated by the flavour puzzle and the Higgs hierarchy problem. While our analysis will mostly focus on BSM flavour models, we stress that this discussion applies more generally to any model featuring the low-scale breaking of semi-simple gauge embeddings of all or part of the $\UIem$ gauge symmetry.

\colorlet{colzoom}{gray!50!NavyBlue}
\begin{figure}[t]\centering\hspace{-1.3em}
\begin{tikzpicture}[declare function = 
  {l1=1.21;r1=9.63;l2=14.42;r2=19.62;
   b1=2.3;t1=18.8;b2=13.85;t2=17.8;}]
% Include the image in a node
% this opacity=0 is a placeholder; I re-load figure at the end
\node [above right,inner sep=0,opacity=0] (image) at (0,0) {%
\raisebox{-0.5\height}{\includegraphics[height=20em]{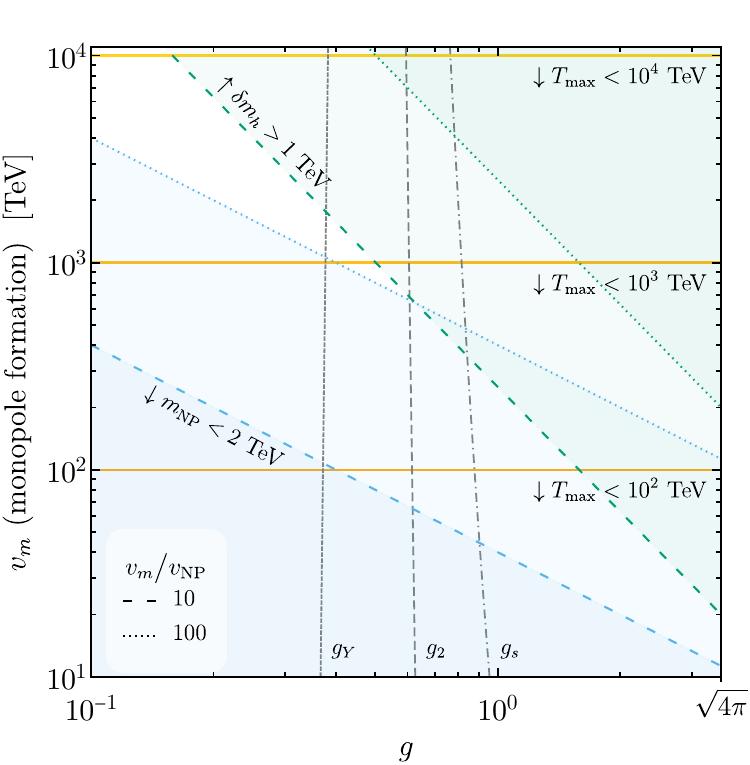}}\hspace*{1.3em}%
\raisebox{-0.5\height}{\includegraphics[height=18em]{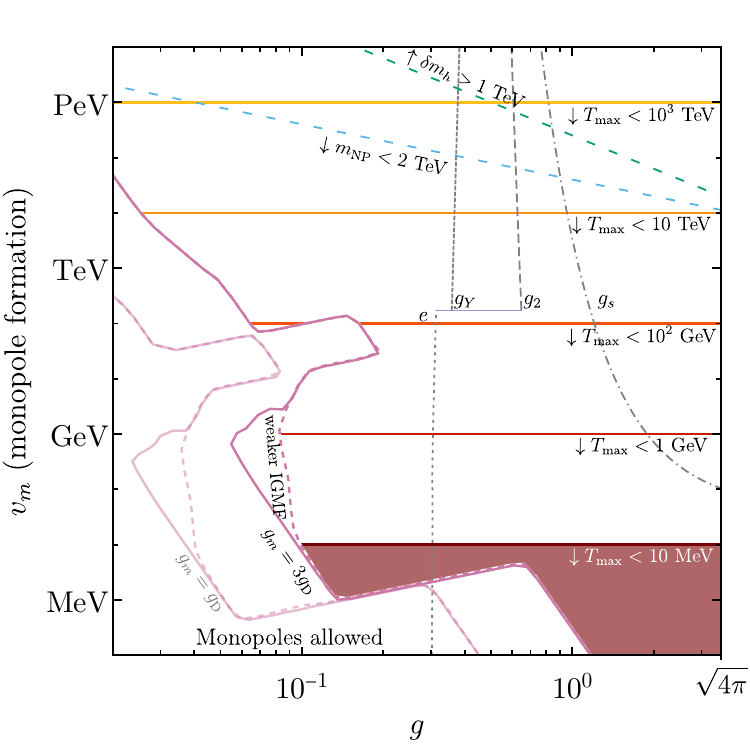}}%
};
% Create scope with normalized axes
\begin{scope}[
x={($0.05*(image.south east)$)},
y={($0.05*(image.north west)$)}]
%% Grid and Axes' labels
%\draw[lightgray,step=1] (image.south west) grid (image.north east);
%\foreach \x in {0,1,...,20} { \node [below] at (\x,0) {\x}; }
%\foreach \y in {0,1,...,20} { \node [left] at (0,\y) {\y};}
%% Labels
%%\node[white,fill=White,opacity=1]
%rectangle for the zoom
\draw[colzoom,opacity=.8,fill=colzoom, fill opacity=.04] (r2,b2) -- (r2,t2) -- (l2,t2) -- (l2,b2) -- cycle;%rectangle
\begin{scope}%Clip segments to the white space above plots
  \clip (r1,t2) rectangle (r2,t1);
  \draw[colzoom,opacity=1] (l1,t1) -- (l2,t2);%NW
  \draw[colzoom,opacity=1] (r1,t1) -- (r2,t2);%NE
\end{scope}
\begin{scope}%Clip segments to the white space between plots
  \clip (r1,b1) rectangle (12.1,t1);
  \draw[colzoom,opacity=1] (r1,b1) -- (r2,b2);%SE
  \draw[colzoom,opacity=1] (l1,b1) -- (l2,b2);%SW
\end{scope}
%\node[above,inner sep=2pt] at (15,0.5) {Text};
%\node[align=center] at (15,18) {Post-infl.~prod.\\ at preheating};
%\node[fill=Goldenrod!70!Dandelion, rounded corners=3mm, fill opacity=0.5, text opacity=1] at (3.5,13) {$\omega$};
\end{scope} 
%re-load the figure, for full visibility
\node [above right,inner sep=0] (image) at (0,0) {%
\raisebox{-0.5\height}{\includegraphics[height=20em]{figures/summary_plot_zoom.pdf}}\hspace*{1.3em}%
\raisebox{-0.5\height}{\includegraphics[height=18em]{figures/summary_plot_wide.pdf}}%
};
\end{tikzpicture}
\vspace{-.5em}
\caption{Summary of collider and cosmological constraints in the $(g,v_m)$ plane, where $g$ denotes the gauge coupling and $v_m$ the symmetry-breaking scale at which monopoles are produced in non-universal flavour completions of the SM. 
The astrophysical and cosmological monopole bounds, shown in pink on the right, require the maximum temperature achieved in the thermal history to satisfy $T_{\max}<v_m$, with representative values indicated by the yellow--orange--brown contours. 
The combination of collider constraints and naturalness considerations motivates the range $v_m\sim 10^{2}$--$10^{4}\,\TeV$, corresponding to the region between the green and blue lines. 
For reference, the grey lines show the values of the SM gauge couplings. 
Two representative choices of the scale separation $v_m/v_\textsc{np}$ are shown with different dashings, as indicated in the legend.
\textit{Left}: region of parameter space relevant for flavour model building. 
In semi-simple gauge extensions involving the third generation, consistency with monopole bounds generically requires $T_{\max}\lesssim 10^{3}$--$10^{4}\,\TeV$. 
\textit{Right}: zoomed-out view of a wider parameter region, extending down to the lowest allowed reheating temperatures, $T_{\max}\gtrsim T_\textsc{BBN}$, shown by the brown shading. 
For the monopole constraints, we show the envelope of the bounds in \cref{fig:monopole astro} (dashed lines correspond to weaker IGMF assumptions, while the lighter shading denotes the case $g_m=\gD$).}
\label{fig:summary plot}
\end{figure}

As highlighted in Sec.~\ref{subsec:FD model examples}, well-motivated flavour models generically predict semi-simple gauge factors whose breaking at a scale $v_m$ gives rise to a residual $\U{1}_{\rm P}$ factor, with $P$ typically corresponding to $B-L$, $R$, or $Y$. This transition, occurring at a temperature $T \lesssim g v_m$, where $g$ denotes the gauge coupling of the parent group containing $\U{1}_{\rm P}$, produces monopoles at scales ranging between $10$ and $10^3\,\mathrm{TeV}$ (see Fig.~\ref{fig:sketch}), many orders of magnitude below the conventional GUT scenario.

Astrophysical and cosmological bounds (pink line in the right panel of \cref{fig:summary plot}, inherited from \cref{fig:monopole astro}) exclude \emph{essentially the entire} parameter space in the absence of monopole dilution. This follows from the combination of the Parker bound, direct searches, and overclosure constraints, which together require the monopole abundance to be exponentially suppressed. The only possible exception lies at parametrically small couplings, $g \ll \mathcal{O}(10^{-2})$, where monopole production and/or annihilation becomes inefficient. However, such values are significantly smaller than the SM gauge couplings and are not expected in realistic semi-simple UV completions with perturbative matching.

Collider constraints (shown as lower limits in blue in \cref{fig:summary plot}, left panel) together with naturalness considerations (green upper boundaries) select a preferred region for new physics signals,
\begin{equation}
v_m \sim 10^{2}\text{--}10^{4}~\mathrm{TeV}, 
\qquad 
g \sim \mathcal{O}(0.1\text{--}1),
\end{equation}
corresponding to the green-blue band in the figure. In this region, the monopole mass lies in the range $m \simeq 4\pi v_m/g \sim 10^{3}\text{--}10^{5}~\text{TeV}$, and the entire band falls deep inside the monopole-excluded region. This demonstrates a direct and unavoidable tension between flavour-motivated constructions and cosmological constraints.

As a consequence, low-scale inflation is not optional: without a period of inflation diluting the monopole abundance, the predicted relic density exceeds observational bounds by many orders of magnitude. Viability of these models requires reheating to occur below the monopole production scale, effectively removing monopoles from the thermal history. Alternative solutions to the monopole problem would also need to take place at scales close to or below the EW scale.

The corresponding maximum temperature $\TRH$ achieved in the thermal history after inflation is shown by the yellow-orange lines in Fig.~\ref{fig:summary plot}. 
This is one of the main messages of this paper: with reasonable assumptions that are common to a wide set of BSM models (particularly in the framework of Flavour Deconstruction), primordial inflation has to occur at relatively low scales, removing more than half of the otherwise allowed orders of magnitude for $\TRH$:
\begin{equation}
\TRH \lesssim v_m \sim 10^{3} \text{--} 10^4\TeV\,.
\end{equation}
In this sense, the production of monopoles has much more drastic and relevant consequences for the cosmological history than in GUT theories, which often lead to monopole production just around the highest $\TRH$ allowed by the bound on primordial tensor modes.  

We emphasise that flavour-motivated gauge extensions probe and constrain the early Universe and its thermal history in a qualitatively different regime compared to GUTs.
The right panel of \cref{fig:summary plot} emphasizes that the upper limit on $\TRH$ implied by flavour is many orders of magnitude closer to the lower limit implied by BBN, than in GUT theories. The resulting constraints from monopole production are therefore significantly stronger for this class of flavour theories.
It is also worth noting that models like e.g.~Froggatt-Nielsen featuring the breaking of semi-simple groups to $\U{1}$ that do not overlap with $\U{1}_{\rm EM}$ are subject to the overabundance bound, which is far weaker than astrophysical constraints but still relevant for $v_m\gtrsim 10^{3}$--$10^5$ TeV.

In the flavour-deconstructed setup which we focus on, the dependence on the hierarchy $v_m/v_{\mathrm{NP}}$ is mild: varying this ratio (e.g.~between $\mathcal{O}(10)$ and $\mathcal{O}(100)$ as shown) only shifts slightly the preferred region in \cref{fig:summary plot}. This conclusion is not an artefact of a specific scale separation, but a generic feature of the framework.
The result is also robust against uncertainties in monopole dynamics, including annihilation efficiency and assumptions on GMF and IGMF (see also the discussion in Appendix~\ref{app:monopole details} for the different velocity regimes of monopoles in our galaxy). \cref{fig:summary plot} uses the envelope of constraints from \cref{fig:monopole astro}, which illustrates that the monopole problem requires a solution like primordial inflation. 

These results represent a direct and quantitative link between flavour structure and cosmology: the presence of semi-simple flavour gauge factors at $v_m \sim 10^{2}\text{--}10^{4}$~TeV implies that the maximal temperature of the Universe has to lie below the same scale, establishing a direct connection between flavour physics, collider-scale new physics, and the reheating history of the Universe.
Conversely, any future probe of the inflationary scale, for example the detection of a primordial GW background in LISA or higher-frequency detectors showing that no inflation occurred from that epoch until today, can exclude or support entire classes of BSM models.
A future detection of a primordial background at LISA, mid-band detectors or ET/CE would imply $T_{\max} \gg 10^{4}$~TeV and would therefore rule out semi-simple completions in this range, while a null result can be seen as a motivation (together with the flavour puzzle and the hierarchy problem) for low-scale inflation and for searches of new physics close to the present reach of colliders.

%%%%%%%%%%%%%%%%%%%%%%%%%%%%%%%%
\section{Conclusion}
\label{sec:conclusions}
In this work, we identified a generic and phenomenologically relevant connection between motivated BSM extensions and the cosmological history of the early Universe. In particular, the absence of new physics signals at the LHC, together with the Higgs hierarchy problem and the hierarchical structure of the Yukawa couplings, motivates TeV-scale flavour non-universal dynamics predominantly coupled to the Higgs and third-generation fermions. Such scenarios naturally arise in flavour non-universal models, which predict semi-simple embeddings of the SM gauge symmetries whose sequential breaking at different scales can account for the observed flavour structure while remaining compatible with current collider and precision bounds.

A generic consequence of these constructions is the presence of intermediate symmetry-breaking stages containing a $\U{1}$ factor with non-trivial overlap with $\U{1}_{\rm EM}$, which unavoidably leads to the production of topological magnetic monopoles. While flavour non-universal models provide an explicit realisation of such scenarios, we stress that our results in \cref{fig:monopole astro} apply more generally to any model featuring the low-scale breaking of semi-simple gauge embeddings of the EM gauge symmetry. 
In particular, in contrast to traditional GUT scenarios, the relevant symmetry-breaking scales in flavour non-universal models, intrinsically tied to the generation of Yukawa hierarchies and the suppression of BSM sources of flavour violation, are much lower, yielding comparatively light monopoles subject to stringent cosmological, astrophysical, and direct-search constraints.

Combining these constraints, we find that the parameter region naturally predicted by flavour models to address the flavour puzzle and generate observable new-physics effects, namely $v_m \sim 10$--$10^{3}\,\mathrm{TeV}$ and $g \sim \mathcal{O}(0.1$--$1)$, is essentially excluded unless monopoles are diluted. This conclusion is robust against uncertainties in the monopole production and dynamics, and reflects the fact that magnetic monopoles produced at these scales are too abundant to be compatible with observations. The viability of this broad class of flavour models, and more generally of any low-scale semi-simple embedding of $\UIem$, therefore requires inflation to dilute the monopole abundance, followed by reheating below the monopole-production scale. Other mechanisms may suppress the monopole abundance, at the price of introducing additional assumptions, but inflation is already strongly motivated by a variety of cosmological observations. In the context of flavour deconstruction, this implies an upper bound on the reheating temperature, $T_{\max} \lesssim 10^{3}$--$10^{4}\,\mathrm{TeV}$, with very interesting implications for the cosmology of these models.

Our results thus expose a direct, two-sided interplay between BSM physics and the inflationary history of the Universe: semi-simple groups at multi-TeV scales require a low-scale reheating, while independent information on inflation or reheating can directly constrain the structure of viable UV completions, and specifically flavour models. In particular, a primordial gravitational-wave signal in CMB $B$-mode searches, such as LiteBIRD~\cite{LiteBIRD:2022cnt,Matsumura:2013aja,Hazumi:2019lys}, CMB-S4~\cite{CMB-S4:2020lpa,CMB-S4:2016ple,Abazajian:2019eic}, Simons Observatory~\cite{Ade:2018sbj,SimonsObservatory:2018koc,SimonsObservatory:2019qwx}, or BICEP/Keck~\cite{BICEP:2021xfz,BICEP2:2014owc}, or in stochastic gravitational-wave searches with pulsar-timing arrays, LISA~\cite{LISA:2017pwj,LISACosmologyWorkingGroup:2022jok}, Einstein Telescope (ET)~\cite{ET:2025xjr,Punturo:2010zz,Hild:2010id}, Cosmic Explorer (CE)~\cite{Reitze:2019iox} and intermediate-frequency proposals \cite{Graham:2017pmn,MAGIS-100:2021etm,Graham:2012sy,Badurina:2019hst,Kawamura:2011zz,Crowder:2005nr,AEDGE:2019nxb}, pointing to reheating temperatures well above $\mathcal{O}(10^{4})\,\mathrm{TeV}$, would either exclude semi-simple completions in this range or require highly non-minimal mechanisms capable of efficiently diluting or annihilating monopoles. Conversely, evidence for non-universal gauge dynamics, or more generally for semi-simple gauge embeddings, at multi-TeV scales, from the LHC, precision flavour experiments, or future colliders would strongly point towards a low-scale reheating history.

The viability of broad classes of semi-simple constructions is therefore intrinsically tied to the inflationary history of the Universe, and the observed absence of magnetic monopoles can act as a cosmological probe of a wide class of BSM models, and in particular of flavour-motivated UV completions.

%%%%%%%%%%%%%%%%%%%%%%%%%%%%%%%%
\acknowledgments
We thank Gino Isidori and Takeshi Kobayashi for useful discussions, and Stefan Antusch, Joe Davighi, Admir Greljo, and Gino Isidori for comments on the manuscript. We thank in particular Joe Davighi for insightful suggestions about part of Appendix~A.\\
\begin{small}
M.P. is grateful for the funding from the UZH Candoc grant and the Swiss National Science Foundation (SNSF) through the grant TMSGI2-225951.\\
D.R.~is supported by the ``Rita Levi-Montalcini'' programme for young researchers of MUR, and was supported at U.~of Zurich by the UZH Postdoc Grant 2023 Nr.\,FK-23-130.
\end{small}

%%%%%%%%%%%%%%%%%%%%%%%%%%%%%%%%
%\clearpage
\appendix
\newpage
\section{Condition for the existence of monopoles} 
\label{app:proof}
In this appendix, we show that the vacuum manifold associated with $G\to H$ admits mo\-no\-po\-les\footnote{Similar claims can already be found e.g.~in \cite{Kibble:1980mv}. Here, however, we prove this conclusion in full generality, see the last paragraph of this appendix for further remarks.} whenever $G$ is compact and semi-simple, and the unbroken
group $H$ has Lie algebra $\mathfrak h=\mathfrak k\oplus \mathfrak u(1)_{\rm P}$ with $\mathfrak k$ semi-simple. In the following, we will use $\times$ to denote a global product of groups, while in the main text, it should be understood as specifying only the Lie algebra rather than the global structure of the group. Note that $H$ can differ from the
direct product $H^{\prime}\equiv K\times \U{1}_{\rm P}$ by a quotient, namely $H\cong H^\prime/\Gamma$ where $\Gamma$ is a finite subgroup of the centre of $H^\prime$. 

\paragraph{The case $H=\U1_{\rm P}$} Let us first consider the simpler case $H=\U1_{\rm P}$ such that $\pi_1(H)\cong\mathbb Z$. Our starting point is the following segment of the homotopy exact sequence
associated with the fibration $H\rightarrow G\rightarrow G/H$ (see e.g. Ref. \cite{Naber:1997yu}):
\begin{equation}
    \cdots
    \rightarrow
    \pi_2(G)
    \rightarrow
    \pi_2(G/H)
    \overset{p}{\rightarrow}
    \pi_1(H)
    \overset{q}{\rightarrow}
    \pi_1(G)
    \rightarrow
    \cdots \,,
    \label{eq:les}
\end{equation}
where every map is a group homomorphism and exactness of the sequence means that the image of one homomorphism therein equals the kernel of the next. 

Now, any Lie group $G$ has $\pi_2(G)=I$, which implies that $p$ is injective and thus $\pi_2(G/H) \cong \operatorname{Im}p \cong \operatorname{Ker}q$ due to the exactness of the sequence.
To determine $\operatorname{Ker}q$, notice that since $\pi_1(H)\cong\mathbb Z$, the homomorphism $q$ is entirely fixed by the image of the generator
of $\pi_1(H)$, which we denote as $\gamma$.
By Weyl's theorem, since $G$ is a compact semi-simple Lie group, $\pi_1(G)$ must be finite. 
Consequently, there must be a smallest positive integer $n$ such that $nf=\mathrm{id}\in {\pi_1(G)}$, where $f$ is the image of $\gamma$ under $q$, i.e.\ $f\equiv q(\gamma)\in \pi_1(G)$.
It follows that $\operatorname{Ker}q$ is an integer subgroup of $\pi_1(H)$ generated by $n\gamma$. Hence, one obtains
\begin{equation}
    \pi_2(G/H)
    \cong
    \mathbb Z \Rightarrow \pi_2(G/H)\neq I \,,
\end{equation}
and this symmetry-breaking pattern leads to the formation of monopoles.

\paragraph{The general case}
Before we turn to the general case $H\cong H^\prime/\Gamma$, let us recall that a finitely generated Abelian group is a group of the form $\mathbb{Z}^r \times T$, where $\mathbb{Z}^r$ is the free part with the non-negative integer $r$ being the free rank, and the torsion part $T$ is a product of a finite number of cyclic groups. The relevant homotopy groups appearing in \cref{eq:les}, as well as those below in \cref{eq:another_les}, are all finitely generated Abelian groups (see e.g. Ref. \cite{Mimura:2000xxx}).

Now, for the general case\footnote{We thank Joe Davighi for giving us insightful input about this case.}, consider the following segment of the homotopy exact sequence associated with the fibration $\Gamma\rightarrow H^\prime\rightarrow H$:
\begin{equation}
    \cdots
    \rightarrow
    \pi_1(\Gamma)
    \rightarrow
    \pi_1(H^\prime)
    \xrightarrow{s}
    \pi_1(H)
    \xrightarrow{t}
    \pi_0(\Gamma)
    \rightarrow
    \cdots\,.
    \label{eq:another_les}
\end{equation}
 Here, $\pi_1(H^\prime)\cong \pi_1(K)\times \mathbb Z$ where $\pi_1(K)$ is finite since $K$ is compact and semi-simple.
 Moreover, $\pi_1(\Gamma)=I$ since $\Gamma$ is discrete, so that $s$ is injective. 
 Therefore, $\operatorname{Im}s\cong\pi_1(H^\prime)$, which then implies that $\pi_1(H)$ must have a free rank $r \geq 1$. However, the case $r>1$ is not possible. This can be shown by noting that $\pi_0(\Gamma) \cong \Gamma$ is finite, which implies that $\operatorname{Ker} t$ has to contain $\mathbb{Z}^r$ as a subgroup. Consequently, any $r>1$ would contradict the equality $\operatorname{Ker} t \cong \operatorname{Im}s\cong \pi_1(H^\prime)$, so we conclude that $r=1$, equivalently $\operatorname{Free}(\pi_1(H)) \cong \mathbb{Z}$ where $\operatorname{Free}(\cdot)$ denotes the free part.

We can now return to the computation of $\pi_2(G/H)$. Since $q$ maps $\pi_1(H)$, which has free rank one, to a finite $\pi_1(G)$, $\operatorname{Ker}q$ must also have free rank one. Using again $\operatorname{Ker}q\cong\pi_2(G/H)$, we obtain
\begin{equation}
    \operatorname{Free}\!\left(\pi_2(G/H)\right)
    \cong
    \mathbb Z \Rightarrow \pi_2(G/H)\neq I \,,
\end{equation}
and the vacuum manifold admits monopoles. Of course, the full group $\pi_2(G/H)$ may also contain finite torsion factors, which classify additional monopole sectors. However, this does not affect the existence argument above.
\paragraph{Comment on universal covers} Often in the literature, one may instead consider the universal cover $\widetilde G$
of $G$, for which $\pi_1(\widetilde G)=I$. In that case, the long exact sequence \cref{eq:les} implies $\pi_2(\widetilde G/\widetilde H)
    \cong
    \pi_1(\widetilde H)$, where $\widetilde H$ denotes the preimage of $H$ in
$\widetilde G$. However, determining $\widetilde H$ is usually not straightforward. 
The argument presented above avoids this complication and establishes the existence of monopoles directly for $G/H$.

\section{Monopole velocity regimes}
\label{app:monopole details}
\begin{figure}[t]
    \centering
    \includegraphics[width=0.49\textwidth]{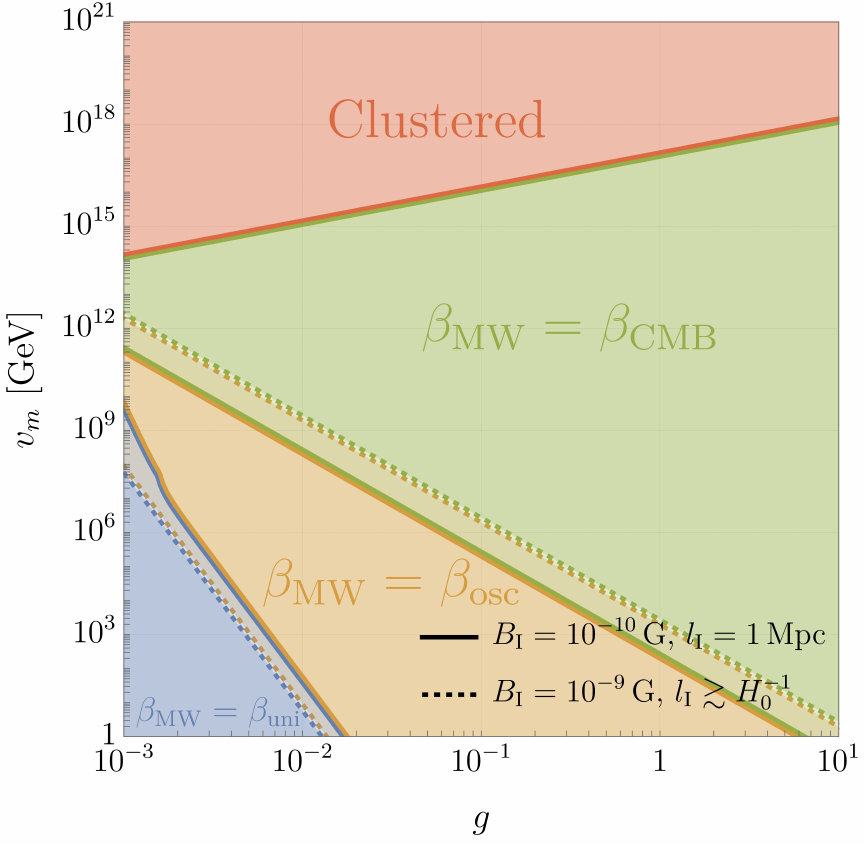}
    \includegraphics[width=0.49\textwidth]{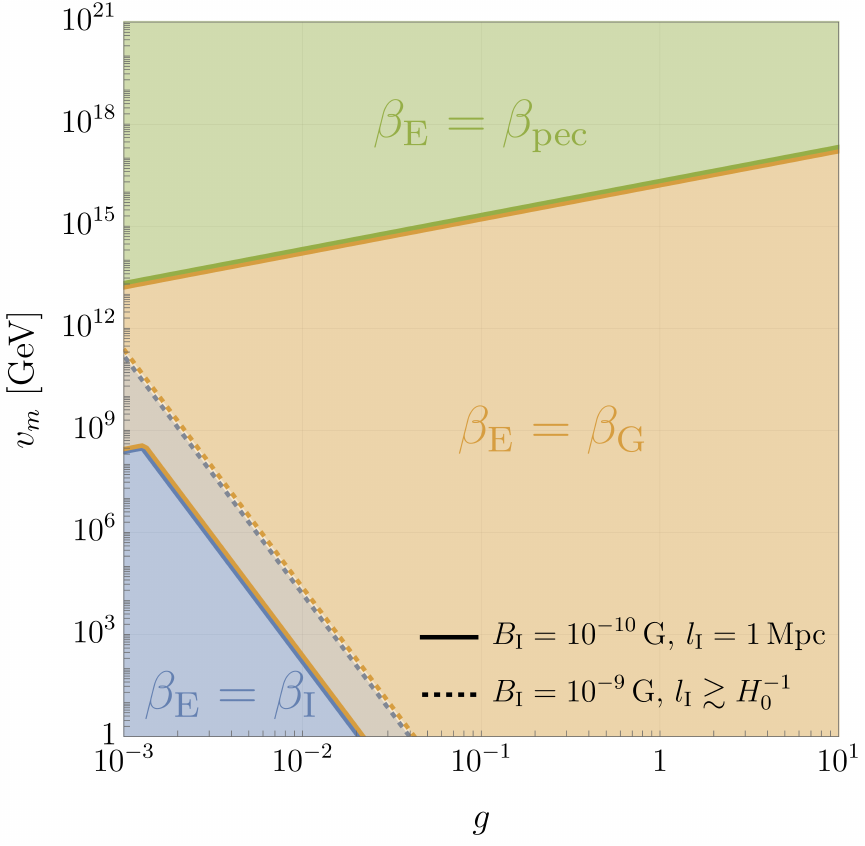}
    \caption{Monopole velocity regimes upon entering the Milky Way (left) and at Earth (right) in the $(g,v_m)$ plane. Solid and dashed contours correspond respectively to the benchmark choices $B_{\rm I}=10^{-10}\,\mathrm{G}$ with coherence length $l_{\rm I}=1\,\mathrm{Mpc}$, and $B_{\rm I}=10^{-9}\,\mathrm{G}$ with $l_{\rm I}\gtrsim H_0^{-1}$. \textit{Left:} Before entering the Galaxy, the velocity of unclustered monopoles is predominantly set either by the peculiar velocity of the Milky Way or by acceleration in the IGMF through magnetic Langmuir oscillations, whereas the uniform acceleration regime is relevant only for sufficiently small vevs and gauge couplings. Heavier monopoles instead cluster gravitationally within the Galaxy. \textit{Right:} Over most of the parameter space, the monopole velocity at Earth is dominated by acceleration in the GMF, while acceleration from the IGMF becomes relevant only for sufficiently light monopoles and small gauge couplings.}
    \label{fig_velregime}
\end{figure}
An important element in this paper is the estimation of the monopole flux in different regimes, generally using the relation $F=n_{\rm cos}\beta/4\pi$, where $n_{\rm cos}$ is the present-day cosmic monopole number density and $\beta$ is the relevant monopole velocity. This allows us to translate the monopole flux bounds of \cite{Perri:2025qpg} into constraints on the parameters of direct interest for model building, namely $g$ and $v_m$, as shown in \cref{fig:monopole astro}. To this end, we need to estimate the monopole velocity, for which we rely mainly on the results of \cite{Perri:2023ncd}, which we briefly summarise in this appendix.

\paragraph{Galactic monopole velocity}
For the Parker bound, if monopoles are sufficiently light that they do not cluster within the Galaxy (roughly $m\lesssim 10^{18}\,\GeV$~\cite{Kobayashi:2023ryr}), the relevant velocity is that of monopoles entering the Milky Way, $\beta_{\mathrm{MW}}$. This velocity can be estimated as
\begin{equation}
\beta_{\mathrm{MW}} \sim \max \left(\beta_{\mathrm{I}}, \beta_{\mathrm{CMB}}\right),
\label{eq:betaMW}
\end{equation}
where $\beta_{\mathrm{I}}$ denotes the monopole velocity induced by acceleration in IGMFs, and $\beta_{\mathrm{CMB}}\approx 10^{-3}$ is the peculiar velocity of the Milky Way in the CMB frame. The monopole velocity induced by IGMF acceleration depends on whether the backreaction of monopoles on the IGMF is negligible. In particular,
\begin{equation}
\beta_{\mathrm{I}} \sim \min \left(\beta_{\mathrm{uni}}, \beta_{\mathrm{osc}}\right).
\label{eq:defbetaI}
\end{equation}
The regime $\beta_{\rm I}\sim\beta_{\mathrm{uni}}$ corresponds to negligible backreaction, so that monopoles are accelerated uniformly within each coherent IGMF cell. In contrast, the regime $\beta_{\rm I}\sim\beta_{\mathrm{osc}}$ corresponds to large backreaction, in which case monopoles undergo the magnetic analogue of Langmuir oscillations. The expressions for $\beta_{\mathrm{uni}}$ and $\beta_{\mathrm{osc}}$ can be found in Ref. \cite{Perri:2023ncd}. These quantities depend on the monopole mass $m$ and magnetic charge $\gem$, the IGMF configuration, namely the typical field strength $B_{\rm I}$ and coherence length $l_{\rm I}$, and the cosmic monopole number density $n_{\rm cos}$. The different regimes for $\beta_{\rm MW}$ in the $(g,v_m)$ plane are shown in the left panel of Fig.~\ref{fig_velregime}. In plotting Fig.~\ref{fig_velregime}, we use the same benchmark parameters as in Fig.~\ref{fig:monopole astro}.

In the regime in which monopoles cluster within the Galaxy, their local number density is no longer given by the cosmic value. In this case, the monopole flux can be estimated as
\begin{equation}
F_{\text{clustered}}\simeq \frac{n_{\text{loc}}\beta_{\mathrm{vir}}}{4\pi},
\end{equation}
where $n_{\text{loc}}$ is the local monopole number density in the Galaxy and $\beta_{\mathrm{vir}}\approx 10^{-3}$ is the Galactic virial velocity. We assume that monopoles cluster within the Galaxy in the same way as dark matter (DM), such that
\begin{equation}
n_{\text{loc}}\simeq \frac{\rho_{\text{DM,loc}}}{\rho_{\text{DM,cos}}}\,n_{\rm cos},
\end{equation}
where $\rho_{\text{DM,loc}}\approx 0.3\GeV\,\mathrm{cm}^{-3}$ and $\rho_{\text{DM,cos}}\approx 1.3\times 10^{-6}\GeV\,\mathrm{cm}^{-3}$ are the DM energy densities in the Solar neighbourhood and in the Universe \cite{ParticleDataGroup:2024cfk}. 
As shown in \cref{fig_velregime}, this regime is mainly relevant for very heavy monopoles, which are typically not produced by spontaneous symmetry breaking of semi-simple gauge groups in flavour models. Moreover, avoiding the overclosure of the Universe generically excludes these monopoles (see \cref{fig:monopole astro}).

\paragraph{Monopole velocity at Earth}
We now turn to the monopole velocity relevant for detection on Earth, $\beta_{\rm E}$. Before reaching Earth, monopoles can be further accelerated by the GMF, acquiring a kinetic energy \cite{Kobayashi:2023ryr}
\begin{equation}
m\left(\gamma_{\mathrm{G}}-1\right)
\sim \gem B_{\mathrm{G}}\sqrt{l_{\mathrm{G}}R}
\simeq 10^{11}\GeV\left(\frac{\gem}{\gemD}\right),
\label{eq:betaG}
\end{equation}
where $\gamma_{\rm G}$ is the monopole Lorentz factor, $B_{\mathrm{G}}$ and $l_{\mathrm{G}}$ denote the strength and coherence length of the GMF, and $R$ is the size of the Galaxy. Note that the GMF-induced velocity $\beta_{\rm G}$ is defined in the Galactic frame, whereas $\beta_{\mathrm{I}}$ is defined in the CMB rest frame. The monopole velocity at Earth can then be estimated as
\begin{equation}
\beta_{\mathrm{E}}
\sim \max \left(\beta_{\mathrm{I}},\beta_{\mathrm{G}},\beta_{\mathrm{CMB}}\right).
\label{eq:betE}
\end{equation}
In writing this expression, it is implicitly assumed that for very heavy monopoles, $\beta_{\mathrm{E}}$ is generally set by the Galactic motion, $\beta_{\mathrm{CMB}}$, and we thus neglect the case in which monopoles cluster within the Galaxy. This does not affect our results, since clustered monopoles are sufficiently heavy that they lie outside the current detection range (see Ref. \cite{Perri:2025qpg}). Moreover, they are generically excluded by monopole overclosure constraints (see \cref{fig:monopole astro}). Consequently, the cosmic monopole number density can still be used to estimate the monopole flux at Earth without accounting for clustering.

For relatively light monopoles, the velocity on Earth can be dominated either by $\beta_{\mathrm{I}}$ or by $\beta_{\mathrm{G}}$. For the IGMF-induced contribution to dominate, namely $\beta_{\mathrm{I}}\gtrsim \beta_{\mathrm{G}}$, one requires both $\beta_{\mathrm{uni}}\gtrsim \beta_{\mathrm{G}}$ and $\beta_{\mathrm{osc}}\gtrsim \beta_{\mathrm{G}}$. Assuming that monopoles are sufficiently light for their terminal velocity to be relativistic, these conditions imply, respectively,
\begin{align}
B_{\mathrm{I}}
&\gtrsim B_{\mathrm{G}}
\sqrt{\frac{l_{\mathrm{G}}R H_0^2}{\min\left(1,l_{\mathrm{I}}H_0\right)}},
\qquad
F_{\mathrm{I}}
\lesssim
\frac{B_{\mathrm{I}}^2}{8\pi \gem B_{\mathrm{G}}\left(l_{\mathrm{G}}R\right)^{1/2}},
\label{eq:cond2bis}
\end{align}
where $F_{\rm I}\sim n_{\rm cos}/4\pi$ is the monopole flux within the IGMF. While the first condition in \cref{eq:cond2bis} can be easily satisfied, the second cannot be satisfied over a wide region of the $(g,v_m)$ parameter space because of the large monopole number density, as illustrated in the right panel of Fig.~\ref{fig_velregime}. Therefore, if a relativistic monopole flux is ever observed on Earth, it should generally be expected to arise from acceleration by the GMF.
\newpage

\bibliographystyle{JHEP}
\bibliography{ref}

@article{Fabri:2025fsc,
    author = "Fabri, Noemi and Isidori, Gino and Racco, Davide",
    title = "{Probing Flavour Deconstruction via Primordial Gravitational Waves}",
    eprint = "2509.12414",
    journal = "accepted for publication in Eur.\ Phys.\ J.\ Plus",
    archivePrefix = "arXiv",
    primaryClass = "hep-ph",
    reportNumber = "ZU-TH 55/25",
    month = "9",
    year = "2025"
}

@article{Lazarides:2024niy,
    author = "Lazarides, George and Maji, Rinku and Shafi, Qaisar",
    title = "{Quantum tunneling in the early universe: stable magnetic monopoles from metastable cosmic strings}",
    eprint = "2402.03128",
    archivePrefix = "arXiv",
    primaryClass = "hep-ph",
    doi = "10.1088/1475-7516/2024/05/128",
    journal = "JCAP",
    volume = "05",
    pages = "128",
    year = "2024"
}

@article{Davighi:2022vpl,
    author = "Davighi, Joe",
    title = "{Gauge flavour unification: From the flavour puzzle to stable protons}",
    eprint = "2206.04482",
    archivePrefix = "arXiv",
    primaryClass = "hep-ph",
    doi = "10.1393/ncc/i2023-23023-0",
    journal = "Nuovo Cim. C",
    volume = "46",
    number = "1",
    pages = "23",
    year = "2022"
}

@article{Mavromatos:2020gwk,
    author = "Mavromatos, Nick E. and Mitsou, Vasiliki A.",
    title = "{Magnetic monopoles revisited: Models and searches at colliders and in the Cosmos}",
    eprint = "2005.05100",
    archivePrefix = "arXiv",
    primaryClass = "hep-ph",
    reportNumber = "KCL-PH-TH/2019-94, IFIC/20-12",
    doi = "10.1142/S0217751X20300124",
    journal = "Int. J. Mod. Phys. A",
    volume = "35",
    number = "23",
    pages = "2030012",
    year = "2020"
}

@article{Mitsou:2026zcf,
    author = "Mitsou, Vasiliki A.",
    title = "{Magnetic Monopoles -- From Dirac to the Large Hadron Collider}",
    eprint = "2605.02869",
    archivePrefix = "arXiv",
    primaryClass = "hep-ex",
    month = "5",
    year = "2026"
}

@article{Chrysostomou:2025vrg,
    author = "Chrysostomou, Anna and Cornell, Alan S. and Darm{\'e}, Luc and Deandrea, Aldo and Demartini, Thibault",
    title = "{Gravitational waves from flavoured SU(2) early-universe phase transitions}",
    eprint = "2512.02148",
    archivePrefix = "arXiv",
    primaryClass = "hep-ph",
    month = "12",
    year = "2025"
}

@article{Davighi:2025cqx,
    author = "Davighi, Joe and Isidori, Gino",
    title = "{A Composite Theory of Higgs and Flavour}",
    eprint = "2512.19650",
    archivePrefix = "arXiv",
    primaryClass = "hep-ph",
    reportNumber = "CERN-TH-2025-267, ZU-TH 88/25",
    month = "12",
    year = "2025"
}

@article{Antusch:2026msw,
    author = "Antusch, Stefan and de Medeiros Varzielas, Ivo and Levy, Miguel",
    title = "{Gravity tidings from domain walls: Flavour hierarchies are making waves}",
    eprint = "2603.23395",
    archivePrefix = "arXiv",
    primaryClass = "hep-ph",
    month = "3",
    year = "2026"
}

@article{Brummer:2025inh,
    author = {Br{\"u}mmer, Felix and Ferrante, Giacomo and Fischer, Th{\'e}odore and Frigerio, Michele},
    title = "{No room for minimal monopole dark matter}",
    eprint = "2509.21924",
    archivePrefix = "arXiv",
    primaryClass = "hep-ph",
    doi = "10.1103/bttg-zfr3",
    journal = "Phys. Rev. D",
    volume = "113",
    number = "9",
    pages = "L091701",
    year = "2026"
}

@article{Perri:2025vtn,
    author = "Perri, Daniele",
    title = "{Noninflationary solution to the monopole problem}",
    eprint = "2509.23726",
    archivePrefix = "arXiv",
    primaryClass = "hep-ph",
    doi = "10.1103/zcp4-pkbw",
    journal = "Phys. Rev. D",
    volume = "113",
    number = "2",
    pages = "023547",
    year = "2026"
}

@article{Csaki:2021ozp,
    author = "Cs{\'a}ki, Csaba and Shirman, Yuri and Telem, Ofri and Terning, John",
    title = "{Pairwise Multiparticle States and the Monopole Unitarity Puzzle}",
    eprint = "2109.01145",
    archivePrefix = "arXiv",
    primaryClass = "hep-th",
    doi = "10.1103/PhysRevLett.129.181601",
    journal = "Phys. Rev. Lett.",
    volume = "129",
    number = "18",
    pages = "181601",
    year = "2022"
}

@article{Kibble:1980mv,
    author = "Kibble, T. W. B.",
    title = "{Some Implications of a Cosmological Phase Transition}",
    reportNumber = "ICTP-79-80-23",
    doi = "10.1016/0370-1573(80)90091-5",
    journal = "Phys. Rept.",
    volume = "67",
    pages = "183",
    year = "1980"
}

@article{Dokos:1979vu,
    author = "Dokos, Constantine P. and Tomaras, Theodore N.",
    title = "{Monopoles and Dyons in the SU(5) Model}",
    reportNumber = "HUTP-79/A071",
    doi = "10.1103/PhysRevD.21.2940",
    journal = "Phys. Rev. D",
    volume = "21",
    pages = "2940",
    year = "1980"
}

@article{Linde:1980tu,
    author = "Linde, Andrei D.",
    title = "{Confinement of Monopoles at High Temperatures: A Solution of the Primordial Monopole Problem}",
    reportNumber = "LEBEDEV-80-125, BI-TP-80-20",
    doi = "10.1016/0370-2693(80)90770-4",
    journal = "Phys. Lett. B",
    volume = "96",
    pages = "293--296",
    year = "1980"
}

@article{Dvali:1995cj,
    author = "Dvali, G. R. and Melfo, Alejandra and Senjanovic, Goran",
    title = "{Is There a monopole problem?}",
    eprint = "hep-ph/9507230",
    archivePrefix = "arXiv",
    reportNumber = "IC-95-145, SISSA-81-95-A",
    doi = "10.1103/PhysRevLett.75.4559",
    journal = "Phys. Rev. Lett.",
    volume = "75",
    pages = "4559--4562",
    year = "1995"
}

@article{LiteBIRD:2022cnt,
    author = "Allys, E. and others",
    collaboration = "LiteBIRD",
    title = "{Probing Cosmic Inflation with the LiteBIRD Cosmic Microwave Background Polarization Survey}",
    eprint = "2202.02773",
    archivePrefix = "arXiv",
    primaryClass = "astro-ph.IM",
    doi = "10.1093/ptep/ptac150",
    journal = "PTEP",
    volume = "2023",
    number = "4",
    pages = "042F01",
    year = "2023"
}

@article{Matsumura:2013aja,
    author = "Matsumura, T. and others",
    title = "{Mission design of LiteBIRD}",
    eprint = "1311.2847",
    archivePrefix = "arXiv",
    primaryClass = "astro-ph.IM",
    doi = "10.1007/s10909-013-0996-1",
    journal = "J. Low Temp. Phys.",
    volume = "176",
    pages = "733",
    year = "2014"
}

@article{Hazumi:2019lys,
    author = "Hazumi, M. and others",
    title = "{LiteBIRD: A Satellite for the Studies of B-Mode Polarization and Inflation from Cosmic Background Radiation Detection}",
    doi = "10.1007/s10909-019-02150-5",
    journal = "J. Low Temp. Phys.",
    volume = "194",
    number = "5-6",
    pages = "443--452",
    year = "2019"
}

@article{SimonsObservatory:2019qwx,
    author = "Abitbol, Maximilian H. and others",
    collaboration = "Simons Observatory",
    title = "{The Simons Observatory: Astro2020 Decadal Project Whitepaper}",
    eprint = "1907.08284",
    archivePrefix = "arXiv",
    primaryClass = "astro-ph.IM",
    journal = "Bull. Am. Astron. Soc.",
    volume = "51",
    pages = "147",
    year = "2019"
}

@article{SimonsObservatory:2018koc,
    author = "Ade, Peter and others",
    collaboration = "Simons Observatory",
    title = "{The Simons Observatory: Science goals and forecasts}",
    eprint = "1808.07445",
    archivePrefix = "arXiv",
    primaryClass = "astro-ph.CO",
    doi = "10.1088/1475-7516/2019/02/056",
    journal = "JCAP",
    volume = "02",
    pages = "056",
    year = "2019"
}

@article{CMB-S4:2020lpa,
    author = "Abazajian, Kevork and others",
    collaboration = "CMB-S4",
    title = "{CMB-S4: Forecasting Constraints on Primordial Gravitational Waves}",
    eprint = "2008.12619",
    archivePrefix = "arXiv",
    primaryClass = "astro-ph.CO",
    reportNumber = "FERMILAB-PUB-20-468-AE-SCD",
    doi = "10.3847/1538-4357/ac1596",
    journal = "Astrophys. J.",
    volume = "926",
    number = "1",
    pages = "54",
    year = "2022"
}

@book{CMB-S4:2016ple,
    author = "Abazajian, Kevork N. and others",
    collaboration = "CMB-S4",
    title = "{CMB-S4 Science Book, First Edition}",
    eprint = "1610.02743",
    archivePrefix = "arXiv",
    primaryClass = "astro-ph.CO",
    reportNumber = "FERMILAB-FN-1024-A-AE",
    doi = "10.2172/1352047",
    month = "10",
    year = "2016"
}

@article{Abazajian:2019eic,
    author = "Abazajian, Kevork and others",
    title = "{CMB-S4 Science Case, Reference Design, and Project Plan}",
    eprint = "1907.04473",
    archivePrefix = "arXiv",
    primaryClass = "astro-ph.IM",
    reportNumber = "FERMILAB-PUB-19-431-AE-SCD",
    month = "7",
    year = "2019"
}

@article{Ade:2018sbj,
    author = "Ade, P. and others",
    collaboration = "Simons Observatory",
    title = "{The Simons Observatory: Science goals and forecasts}",
    eprint = "1808.07445",
    archivePrefix = "arXiv",
    primaryClass = "astro-ph.CO",
    doi = "10.1088/1475-7516/2019/02/056",
    journal = "JCAP",
    volume = "02",
    pages = "056",
    year = "2019"
}

@article{BICEP2:2014owc,
    author = "Ade, P. A. R. and others",
    collaboration = "BICEP2",
    title = "{Detection of $B$-Mode Polarization at Degree Angular Scales by BICEP2}",
    eprint = "1403.3985",
    archivePrefix = "arXiv",
    primaryClass = "astro-ph.CO",
    doi = "10.1103/PhysRevLett.112.241101",
    journal = "Phys. Rev. Lett.",
    volume = "112",
    number = "24",
    pages = "241101",
    year = "2014"
}

@article{BICEP:2021xfz,
    author = "Ade, P. A. R. and others",
    collaboration = "BICEP, Keck",
    title = "{Improved Constraints on Primordial Gravitational Waves using Planck, WMAP, and BICEP/Keck Observations through the 2018 Observing Season}",
    eprint = "2110.00483",
    archivePrefix = "arXiv",
    primaryClass = "astro-ph.CO",
    doi = "10.1103/PhysRevLett.127.151301",
    journal = "Phys. Rev. Lett.",
    volume = "127",
    number = "15",
    pages = "151301",
    year = "2021"
}

@article{Tristram:2021tvh,
    author = "Tristram, M. and others",
    title = "{Improved limits on the tensor-to-scalar ratio using BICEP and Planck data}",
    eprint = "2112.07961",
    archivePrefix = "arXiv",
    primaryClass = "astro-ph.CO",
    doi = "10.1103/PhysRevD.105.083524",
    journal = "Phys. Rev. D",
    volume = "105",
    number = "8",
    pages = "083524",
    year = "2022"
}

@article{Kawamura:2011zz,
    author = "Kawamura, Seiji and others",
    editor = "Buchman, Sasha and Sun, Ke-Xun",
    title = "{The Japanese space gravitational wave antenna: DECIGO}",
    doi = "10.1088/0264-9381/28/9/094011",
    journal = "Class. Quant. Grav.",
    volume = "28",
    pages = "094011",
    year = "2011"
}

@article{Crowder:2005nr,
    author = "Crowder, Jeff and Cornish, Neil J.",
    title = "{Beyond LISA: Exploring future gravitational wave missions}",
    eprint = "gr-qc/0506015",
    archivePrefix = "arXiv",
    doi = "10.1103/PhysRevD.72.083005",
    journal = "Phys. Rev. D",
    volume = "72",
    pages = "083005",
    year = "2005"
}

@article{AEDGE:2019nxb,
    author = "El-Neaj, Yousef Abou and others",
    collaboration = "AEDGE",
    title = "{AEDGE: Atomic Experiment for Dark Matter and Gravity Exploration in Space}",
    eprint = "1908.00802",
    archivePrefix = "arXiv",
    primaryClass = "gr-qc",
    reportNumber = "KCL-PH-TH/2019-65, CERN-TH-2019-126",
    doi = "10.1140/epjqt/s40507-020-0080-0",
    journal = "EPJ Quant. Technol.",
    volume = "7",
    pages = "6",
    year = "2020"
}

@article{LISA:2017pwj,
    author = "Amaro-Seoane, Pau and others",
    collaboration = "LISA",
    title = "{Laser Interferometer Space Antenna}",
    eprint = "1702.00786",
    archivePrefix = "arXiv",
    primaryClass = "astro-ph.IM",
    doi = "10.48550/arXiv.1702.00786",
    year = "2017"
}

@article{Reitze:2019iox,
    author = "Reitze, David and others",
    title = "{Cosmic Explorer: The U.S. Contribution to Gravitational-Wave Astronomy beyond LIGO}",
    eprint = "1907.04833",
    archivePrefix = "arXiv",
    primaryClass = "astro-ph.IM",
    doi = "10.48550/arXiv.1907.04833",
    year = "2019"
}

@article{Isidori:2010kg,
    author = "Isidori, Gino and Nir, Yosef and Perez, Gilad",
    title = "{Flavor Physics Constraints for Physics Beyond the Standard Model}",
    eprint = "1002.0900",
    archivePrefix = "arXiv",
    primaryClass = "hep-ph",
    doi = "10.1146/annurev.nucl.012809.104534",
    journal = "Ann. Rev. Nucl. Part. Sci.",
    volume = "60",
    pages = "355",
    year = "2010"
}

@article{tHooft:1974kcl,
    author = "'t Hooft, Gerard",
    editor = "Taylor, J. C.",
    title = "{Magnetic Monopoles in Unified Gauge Theories}",
    reportNumber = "CERN-TH-1876",
    doi = "10.1016/0550-3213(74)90486-6",
    journal = "Nucl. Phys. B",
    volume = "79",
    pages = "276--284",
    year = "1974"
}

@article{Kolb:1982si,
    author = "Kolb, Edward W. and Colgate, Stirling A. and Harvey, Jeffrey A.",
    title = "{Monopole Catalysis of Nucleon Decay in Neutron Stars}",
    reportNumber = "LA-UR-82-1963",
    doi = "10.1103/PhysRevLett.49.1373",
    journal = "Phys. Rev. Lett.",
    volume = "49",
    pages = "1373",
    year = "1982"
}

@article{Freese:1983hz,
    author = "Freese, Katherine and Turner, Michael S. and Schramm, David N.",
    title = "{Monopole Catalysis of Nucleon Decay in Old Pulsars}",
    reportNumber = "EFI-83-27-CHICAGO",
    doi = "10.1103/PhysRevLett.51.1625",
    journal = "Phys. Rev. Lett.",
    volume = "51",
    pages = "1625",
    year = "1983"
}

@article{Dvali:1997sa,
    author = "Dvali, G. R. and Liu, Hong and Vachaspati, Tanmay",
    title = "{Sweeping away the monopole problem}",
    eprint = "hep-ph/9710301",
    archivePrefix = "arXiv",
    reportNumber = "CWRU-P15-97, IMPERIAL-TP-97-98-3, CERN-TH-97-273",
    doi = "10.1103/PhysRevLett.80.2281",
    journal = "Phys. Rev. Lett.",
    volume = "80",
    pages = "2281--2284",
    year = "1998"
}

@article{Bajc:1998rd,
    author = "Bajc, Borut and Riotto, Antonio and Senjanovic, Goran",
    title = "{R - charge kills monopoles}",
    eprint = "hep-ph/9803438",
    archivePrefix = "arXiv",
    reportNumber = "IJS-TP-98-4, IC-98-33, CERN-TH-98-3, CERN-TH-98-003",
    doi = "10.1142/S0217732398003132",
    journal = "Mod. Phys. Lett. A",
    volume = "13",
    pages = "2955--2964",
    year = "1998"
}

@article{IceCube:2016Monopoles,
  author       = {Aartsen, M.~G. and Abraham, K. and Ackermann, M. and others},
  collaboration= {IceCube},
  title        = {Searches for relativistic magnetic monopoles in {IceCube}},
  journal      = {Eur. Phys. J. C},
  volume       = {76},
  year         = {2016},
  pages        = {133},
  doi          = {10.1140/epjc/s10052-016-3953-8},
  eprint       = {1602.01617},
  archivePrefix= {arXiv},
  primaryClass = {astro-ph.HE}
}

@article{PhysRev.138.B248,
  title = {Total Energy Loss and \ifmmode \check{C}\else \v{C}\fi{}erenkov Emission from Monopoles},
  author = {Tompkins, Donald R.},
  journal = {Phys. Rev.},
  volume = {138},
  issue = {1B},
  pages = {B248--B250},
  numpages = {0},
  year = {1965},
  month = {Apr},
  publisher = {American Physical Society},
  doi = {10.1103/PhysRev.138.B248},
  url = {https://link.aps.org/doi/10.1103/PhysRev.138.B248}
}

@article{Kobayashi:2023ryr,
    author = "Kobayashi, Takeshi and Perri, Daniele",
    title = "{Parker bounds on monopoles with arbitrary charge from galactic and primordial magnetic fields}",
    eprint = "2307.07553",
    archivePrefix = "arXiv",
    primaryClass = "hep-ph",
    doi = "10.1103/PhysRevD.108.083005",
    journal = "Phys. Rev. D",
    volume = "108",
    number = "8",
    pages = "083005",
    year = "2023"
}

@article{Preskill:1979zi,
    author = "Preskill, John",
    title = "{Cosmological Production of Superheavy Magnetic Monopoles}",
    reportNumber = "HUTP-79/A028",
    doi = "10.1103/PhysRevLett.43.1365",
    journal = "Phys. Rev. Lett.",
    volume = "43",
    pages = "1365",
    year = "1979"
}

@article{Greljo:2019xan,
    author = "Greljo, Admir and Opferkuch, Toby and Stefanek, Ben A.",
    title = "{Gravitational Imprints of Flavor Hierarchies}",
    eprint = "1910.02014",
    archivePrefix = "arXiv",
    primaryClass = "hep-ph",
    reportNumber = "CERN-TH-2019-162",
    doi = "10.1103/PhysRevLett.124.171802",
    journal = "Phys. Rev. Lett.",
    volume = "124",
    number = "17",
    pages = "171802",
    year = "2020"
}

@article{Davighi:2022bqf,
  title = {Electroweak-Flavour and Quark-Lepton Unification: A Family Non-Universal Path},
  author = {Davighi, Joe and Isidori, Gino and Pesut, Marko},
  year = {2023},
  month = apr,
  journal = {JHEP},
  volume = {04},
  eprint = {2212.06163},
  primaryclass = {hep-ph},
  pages = {030},
  doi = {10.1007/JHEP04(2023)030},
  archiveprefix = {arXiv}
}

@article{Davighi:2022fer,
  title = {Electroweak Flavour Unification},
  author = {Davighi, Joe and {Tooby-Smith}, Joseph},
  year = {2022},
  month = sep,
  journal = {JHEP},
  volume = {09},
  eprint = {2201.07245},
  primaryclass = {hep-ph},
  pages = {193},
  doi = {10.1007/JHEP09(2022)193},
  archiveprefix = {arXiv}
}

@article{Davighi:2023evx,
  title = {Deconstructed Hypercharge: A Natural Model of Flavour},
  author = {Davighi, Joe and Stefanek, Ben A.},
  year = {2023},
  month = nov,
  journal = {JHEP},
  volume = {11},
  number = {ZU-TH 24/23},
  eprint = {2305.16280},
  primaryclass = {hep-ph},
  pages = {100},
  doi = {10.1007/JHEP11(2023)100},
  archiveprefix = {arXiv}
}

@article{Davighi:2023iks,
  title = {Non-Universal Gauge Interactions Addressing the Inescapable Link between {{Higgs}} and Flavour},
  author = {Davighi, Joe and Isidori, Gino},
  year = {2023},
  month = jul,
  journal = {JHEP},
  volume = {07},
  eprint = {2303.01520},
  primaryclass = {hep-ph},
  pages = {147},
  doi = {10.1007/JHEP07(2023)147},
  archiveprefix = {arXiv}
}

@article{Davighi:2023xqn,
  title = {Phenomenology of a Deconstructed Electroweak Force},
  author = {Davighi, Joe and Gosnay, Alastair and Miller, David J. and Renner, Sophie},
  year = {2024},
  month = may,
  journal = {JHEP},
  volume = {05},
  number = {CERN-TH-2023-246},
  eprint = {2312.13346},
  primaryclass = {hep-ph},
  pages = {085},
  doi = {10.1007/JHEP05(2024)085},
  archiveprefix = {arXiv}
}

@article{Tino:2019tkb,
    author = "Tino, Guglielmo M. and others",
    title = "{SAGE: A Proposal for a Space Atomic Gravity Explorer}",
    eprint = "1907.03867",
    archivePrefix = "arXiv",
    primaryClass = "astro-ph.IM",
    doi = "10.1140/epjd/e2019-100324-6",
    journal = "Eur. Phys. J. D",
    volume = "73",
    number = "11",
    pages = "228",
    year = "2019"
}

@article{Langacker:1980kd,
    author = "Langacker, Paul and Pi, So-Young",
    title = "{Magnetic Monopoles in Grand Unified Theories}",
    reportNumber = "SLAC-PUB-2496, Print-80-0262 (IAS,PRINCETON)",
    doi = "10.1103/PhysRevLett.45.1",
    journal = "Phys. Rev. Lett.",
    volume = "45",
    pages = "1",
    year = "1980"
}

@article{MAGIS-100:2021etm,
    author = "Abe, Mahiro and others",
    collaboration = "MAGIS-100",
    title = "{Matter-wave Atomic Gradiometer Interferometric Sensor (MAGIS-100)}",
    eprint = "2104.02835",
    archivePrefix = "arXiv",
    primaryClass = "physics.atom-ph",
    reportNumber = "FERMILAB-PUB-21-031-AD-DI-FESS-QIS-T",
    doi = "10.1088/2058-9565/abf719",
    journal = "Quantum Sci. Technol.",
    volume = "6",
    number = "4",
    pages = "044003",
    year = "2021"
}

@inproceedings{Davighi:2025icd,
  title = {Higgs and Flavour: {{BSM}} Overview},
  booktitle = {{{35th Rencontres de Blois}}: {{Particle Physics and Cosmology}}},
  author = {Davighi, Joe},
  year = {2025},
  month = jan,
  eprint = {2501.16064},
  primaryclass = {hep-ph},
  archiveprefix = {arXiv}
}

@article{Fritzsch:1974nn,
    author = "Fritzsch, Harald and Minkowski, Peter",
    title = "{Unified Interactions of Leptons and Hadrons}",
    reportNumber = "CALT-68-467",
    doi = "10.1016/0003-4916(75)90211-0",
    journal = "Annals Phys.",
    volume = "93",
    pages = "193--266",
    year = "1975"
}

@article{Dimopoulos:1981yj,
    author = "Dimopoulos, S. and Raby, S. and Wilczek, Frank",
    title = "{Supersymmetry and the Scale of Unification}",
    reportNumber = "NSF-ITP-81-31",
    doi = "10.1103/PhysRevD.24.1681",
    journal = "Phys. Rev. D",
    volume = "24",
    pages = "1681--1683",
    year = "1981"
}

@article{Dimopoulos:1981zb,
    author = "Dimopoulos, Savas and Georgi, Howard",
    title = "{Softly Broken Supersymmetry and SU(5)}",
    reportNumber = "HUTP-81/A022",
    doi = "10.1016/0550-3213(81)90522-8",
    journal = "Nucl. Phys. B",
    volume = "193",
    pages = "150--162",
    year = "1981"
}

@article{Dimopoulos:1982cz,
    author = "Dimopoulos, Savas and Preskill, John and Wilczek, Frank",
    title = "{Catalyzed Nucleon Decay in Neutron Stars}",
    reportNumber = "HUTP-82-A047, NSF-ITP-82-91",
    doi = "10.1016/0370-2693(82)90679-7",
    journal = "Phys. Lett. B",
    volume = "119",
    pages = "320",
    year = "1982"
}

@article{Greljo:2024ovt,
  title = {Neutrino Anarchy from Flavor Deconstruction},
  author = {Greljo, Admir and Isidori, Gino},
  year = {2024},
  month = jul,
  journal = {Physics Letters B},
  volume = {856},
  eprint = {2406.01696},
  primaryclass = {hep-ph},
  pages = {138900},
  doi = {10.1016/j.physletb.2024.138900},
  archiveprefix = {arXiv}
}

@article{Fuentes-Martin:2020pww,
    author = "Fuentes-Martin, Javier and Isidori, Gino and Pag{\`e}s, Julie and Stefanek, Ben A.",
    title = "{Flavor non-universal Pati-Salam unification and neutrino masses}",
    eprint = "2012.10492",
    archivePrefix = "arXiv",
    primaryClass = "hep-ph",
    reportNumber = "MITP/20-083, ZU-TH-56/20",
    doi = "10.1016/j.physletb.2021.136484",
    journal = "Phys. Lett. B",
    volume = "820",
    pages = "136484",
    year = "2021"
}

@article{Pati:1974yy,
    author = "Pati, Jogesh C. and Salam, Abdus",
    title = "{Lepton Number as the Fourth Color}",
    reportNumber = "IC-74-7",
    doi = "10.1103/PhysRevD.10.275",
    journal = "Phys. Rev. D",
    volume = "10",
    pages = "275--289",
    year = "1974",
    note = "[Erratum: Phys.Rev.D 11, 703--703 (1975)]"
}

@article{Faroughy:2020ina,
    author = "Faroughy, Darius A. and Isidori, Gino and Wilsch, Felix and Yamamoto, Kei",
    title = "{Flavour symmetries in the SMEFT}",
    eprint = "2005.05366",
    archivePrefix = "arXiv",
    primaryClass = "hep-ph",
    doi = "10.1007/JHEP08(2020)166",
    journal = "JHEP",
    volume = "08",
    pages = "166",
    year = "2020"
}

@article{Georgi:1974sy,
    author = "Georgi, H. and Glashow, S. L.",
    title = "{Unity of All Elementary Particle Forces}",
    doi = "10.1103/PhysRevLett.32.438",
    journal = "Phys. Rev. Lett.",
    volume = "32",
    pages = "438--441",
    year = "1974"
}

@article{DAgnolo:2012ulg,
    author = "D'Agnolo, Raffaele Tito and Straub, David M.",
    title = "{Gauged flavour symmetry for the light generations}",
    eprint = "1202.4759",
    archivePrefix = "arXiv",
    primaryClass = "hep-ph",
    doi = "10.1007/JHEP05(2012)034",
    journal = "JHEP",
    volume = "05",
    pages = "034",
    year = "2012"
}

@article{Linster:2018avp,
    author = "Linster, Matthias and Ziegler, Robert",
    title = "{A Realistic $U(2)$ Model of Flavor}",
    eprint = "1805.07341",
    archivePrefix = "arXiv",
    primaryClass = "hep-ph",
    reportNumber = "CERN-TH-2018-121, TTP18-019",
    doi = "10.1007/JHEP08(2018)058",
    journal = "JHEP",
    volume = "08",
    pages = "058",
    year = "2018"
}

@article{Greljo:2023bix,
    author = "Greljo, Admir and Thomsen, Anders Eller",
    title = "{Rising through the ranks: flavor hierarchies from a gauged SU(2) symmetry}",
    eprint = "2309.11547",
    archivePrefix = "arXiv",
    primaryClass = "hep-ph",
    doi = "10.1140/epjc/s10052-024-12556-5",
    journal = "Eur. Phys. J. C",
    volume = "84",
    number = "2",
    pages = "213",
    year = "2024"
}

@article{Darme:2023nsy,
    author = "Darm{\'e}, Luc and Deandrea, Aldo and Mahmoudi, Farvah",
    title = "{Gauge SU(2)$_{f}$ flavour transfers}",
    eprint = "2307.09595",
    archivePrefix = "arXiv",
    primaryClass = "hep-ph",
    reportNumber = "CERN-TH-2023-139",
    doi = "10.1007/JHEP05(2024)313",
    journal = "JHEP",
    volume = "05",
    pages = "313",
    year = "2024"
}

@article{Farina:2013mla,
    author = "Farina, Marco and Pappadopulo, Duccio and Strumia, Alessandro",
    title = "{A modified naturalness principle and its experimental tests}",
    eprint = "1303.7244",
    archivePrefix = "arXiv",
    primaryClass = "hep-ph",
    doi = "10.1007/JHEP08(2013)022",
    journal = "JHEP",
    volume = "08",
    pages = "022",
    year = "2013"
}

@article{FernandezNavarro:2023hrf,
    author = "Fern{\'a}ndez Navarro, Mario and King, Stephen F. and Vicente, Avelino",
    title = "{Tri-unification: a separate SU(5) for each fermion family}",
    eprint = "2311.05683",
    archivePrefix = "arXiv",
    primaryClass = "hep-ph",
    reportNumber = "IFIC/23-48",
    doi = "10.1007/JHEP05(2024)130",
    journal = "JHEP",
    volume = "05",
    pages = "130",
    year = "2024"
}

@article{Allanach:2021bfe,
    author = "Allanach, B. C. and Gripaios, Ben and Tooby-Smith, Joseph",
    title = "{Semisimple extensions of the Standard Model gauge algebra}",
    eprint = "2104.14555",
    archivePrefix = "arXiv",
    primaryClass = "hep-th",
    doi = "10.1103/PhysRevD.104.035035",
    journal = "Phys. Rev. D",
    volume = "104",
    number = "3",
    pages = "035035",
    year = "2021",
    note = "[Erratum: Phys.Rev.D 106, 019901 (2022)]"
}

@article{Gosnay:2026dye,
    author = "Gosnay, Alastair and Miller, David J.",
    title = "{$\mathrm{Sp}(6)$ Unifying Deconstructed $\mathrm{SU}(2)$'s}",
    eprint = "2603.27359",
    archivePrefix = "arXiv",
    primaryClass = "hep-ph",
    month = "3",
    year = "2026"
}

@article{Fuentes-Martin:2019ign,
    author = "Fuentes-Martin, Javier and Isidori, Gino and Konig, Matthias and Selimovic, Nikola",
    title = "{Vector Leptoquarks Beyond Tree Level}",
    eprint = "1910.13474",
    archivePrefix = "arXiv",
    primaryClass = "hep-ph",
    doi = "10.1103/PhysRevD.101.035024",
    journal = "Phys. Rev. D",
    volume = "101",
    pages = "035024",
    year = "2020"
}

@article{Fuentes-Martin:2020hvc,
    author = "Fuentes-Martin, Javier and Isidori, Gino and Konig, Matthias and Selimovic, Nikola",
    title = "{Vector Leptoquarks Beyond Tree Level III: Vector-like Fermions and Flavor-Changing Transitions}",
    eprint = "2009.11296",
    archivePrefix = "arXiv",
    primaryClass = "hep-ph",
    doi = "10.1103/PhysRevD.102.115015",
    journal = "Phys. Rev. D",
    volume = "102",
    pages = "115015",
    year = "2020"
}

@article{Faroughy:2016osc,
    author = "Faroughy, Dorsner and Greljo, Admir and Kamenik, Jernej F.",
    title = "{Confronting lepton flavor universality violation in B decays with high-$p_T$ tau lepton searches at LHC}",
    eprint = "1609.07138",
    archivePrefix = "arXiv",
    primaryClass = "hep-ph",
    doi = "10.1016/j.physletb.2017.01.015",
    journal = "Phys. Lett. B",
    volume = "764",
    pages = "126--134",
    year = "2017"
}

@article{Greljo:2025mwj,
    author = "Greljo, Admir and Palavri{\'c}, Ajdin and Stefanek, Ben A.",
    title = "{Minimal Flavor Protection for TeV-scale New Physics}",
    eprint = "2512.04159",
    archivePrefix = "arXiv",
    primaryClass = "hep-ph",
    month = "12",
    year = "2025"
}

@article{Altmannshofer:2024hmr,
    author = "Altmannshofer, Wolfgang and Greljo, Admir",
    title = "{Recent Progress in Flavor Model Building}",
    eprint = "2412.04549",
    archivePrefix = "arXiv",
    primaryClass = "hep-ph",
    doi = "10.1146/annurev-nucl-121423-100950",
    journal = "Ann. Rev. Nucl. Part. Sci.",
    volume = "75",
    number = "1",
    pages = "201--322",
    year = "2025"
}

@article{Gripaios:2014tna,
    author = "Gripaios, Ben and Nardecchia, Marco and Renner, Stefan A.",
    title = "{Composite leptoquarks and anomalies in $B$-meson decays}",
    eprint = "1406.5969",
    archivePrefix = "arXiv",
    primaryClass = "hep-ph",
    doi = "10.1007/JHEP05(2015)006",
    journal = "JHEP",
    volume = "05",
    pages = "006",
    year = "2015"
}

@article{Cornella:2021sby,
    author = "Cornella, Claudia and Faroughy, D. A. and Fuentes-Martin, Javier and Isidori, Gino and Neubert, Matthias",
    title = "{Reading the footprints of the B-meson flavor anomalies}",
    eprint = "2103.16558",
    archivePrefix = "arXiv",
    primaryClass = "hep-ph",
    doi = "10.1007/JHEP08(2021)050",
    journal = "JHEP",
    volume = "08",
    pages = "050",
    year = "2021"
}

@article{Rubakov:1981rg,
    author = "Rubakov, V. A.",
    title = "{Adler-Bell-Jackiw Anomaly and Fermion Number Breaking in the Presence of a Magnetic Monopole}",
    journal = "Nucl. Phys. B",
    volume = "203",
    pages = "311--348",
    year = "1982",
    doi = "10.1016/0550-3213(82)90034-7"
}

@article{Blasi:2024vew,
    author = "Blasi, Simone and Calibbi, Lorenzo and Mariotti, Alberto and Turbang, Kevin",
    title = "{Gravitational waves from cosmic strings in Froggatt-Nielsen flavour models}",
    eprint = "2410.08668",
    archivePrefix = "arXiv",
    primaryClass = "hep-ph",
    reportNumber = "DESY-24-147, DESY--24--147",
    doi = "10.1007/JHEP05(2025)019",
    journal = "JHEP",
    volume = "05",
    pages = "019",
    year = "2025"
}

@article{Antusch:2025xrs,
    author = "Antusch, Stefan and Hinze, Kevin and Saad, Shaikh",
    title = "{Metastable cosmic strings and gravitational waves from flavor symmetry breaking}",
    eprint = "2503.05868",
    archivePrefix = "arXiv",
    primaryClass = "hep-ph",
    doi = "10.1103/528x-qzs3",
    journal = "Phys. Rev. D",
    volume = "112",
    number = "3",
    pages = "035043",
    year = "2025"
}

@article{Cordova:2022qtz,
    author = "Cordova, Clay and Koren, Seth",
    title = "{Higher Flavor Symmetries in the Standard Model}",
    eprint = "2212.13193",
    archivePrefix = "arXiv",
    primaryClass = "hep-ph",
    doi = "10.1002/andp.202300031",
    journal = "Annalen Phys.",
    volume = "535",
    number = "8",
    pages = "2300031",
    year = "2023"
}

@article{Callan:1982au,
    author = "Callan, Curtis G. Jr.",
    title = "{Monopole Catalysis of Baryon Decay}",
    journal = "Nucl. Phys. B",
    volume = "212",
    pages = "391--400",
    year = "1983",
    doi = "10.1016/0550-3213(83)90677-7"
}

@article{Patrizii:2015uea,
    author = "Patrizii, Laura and Spurio, Maurizio",
    title = "{Status of Searches for Magnetic Monopoles}",
    eprint = "1510.07125",
    archivePrefix = "arXiv",
    primaryClass = "hep-ex",
    doi = "10.1142/S0217751X15300162",
    journal = "Int. J. Mod. Phys. A",
    volume = "30",
    pages = "1530012",
    year = "2015"
}

@book{Shnir:2005xx,
    author = "Shnir, Yakov M.",
    title = "{Magnetic Monopoles}",
    publisher = "Springer",
    address = "Berlin, Germany",
    year = "2005",
    doi = "10.1007/3-540-29082-6"
}

@article{Glioti:2024hye,
    author = "Glioti, Alfredo and Rattazzi, Riccardo and Ricci, Lorenzo and Vecchi, Luca",
    title = "{Exploring the flavor symmetry landscape}",
    eprint = "2402.09503",
    archivePrefix = "arXiv",
    primaryClass = "hep-ph",
    doi = "10.21468/SciPostPhys.18.6.201",
    journal = "SciPost Phys.",
    volume = "18",
    number = "6",
    pages = "201",
    year = "2025"
}

@article{Bauer:2015knc,
    author = "Bauer, Martin and Neubert, Matthias",
    title = "{Minimal Leptoquark Explanation for the $R_{D^{(*)}}$, $R_K$, and $(g-2)_\mu$ Anomalies}",
    eprint = "1511.01900",
    archivePrefix = "arXiv",
    primaryClass = "hep-ph",
    doi = "10.1103/PhysRevLett.116.141802",
    journal = "Phys. Rev. Lett.",
    volume = "116",
    pages = "141802",
    year = "2016"
}

@article{Crivellin:2017zlb,
    author = "Crivellin, Andreas and Mueller, D. and Ota, T.",
    title = "{Simultaneous explanation of $R(D^{(*)})$ and $b \to s \mu^+ \mu^-$: the last scalar leptoquarks standing}",
    eprint = "1703.09226",
    archivePrefix = "arXiv",
    primaryClass = "hep-ph",
    doi = "10.1007/JHEP09(2017)040",
    journal = "JHEP",
    volume = "09",
    pages = "040",
    year = "2017"
}

@article{Davighi:2022dyq,
    author = "Davighi, Joe and Tooby-Smith, Joseph",
    title = "{Flatland: abelian extensions of the Standard Model with semi-simple completions}",
    eprint = "2206.11271",
    archivePrefix = "arXiv",
    primaryClass = "hep-ph",
    doi = "10.1007/JHEP09(2022)159",
    journal = "JHEP",
    volume = "09",
    pages = "159",
    year = "2022"
}

@article{Gould:2019myj,
    author = "Gould, Oliver and Ho, David L. -J. and Rajantie, Arttu",
    title = "{Towards Schwinger production of magnetic monopoles in heavy-ion collisions}",
    eprint = "1902.04388",
    archivePrefix = "arXiv",
    primaryClass = "hep-th",
    reportNumber = "IMPERIAL-TP-2019-DH-01, HIP-2019-2/TH, IMPERIAL-TP-2019-DH-01; HIP-2019-2/TH",
    doi = "10.1103/PhysRevD.100.015041",
    journal = "Phys. Rev. D",
    volume = "100",
    number = "1",
    pages = "015041",
    year = "2019"
}

@inproceedings{Pesut:2025ook,
    author = "Pesut, Marko",
    title = "{Flavour Non-Universality and Higgs Compositeness}",
    booktitle = "{35th Rencontres de Blois}: {Particle Physics and Cosmology}",
    eprint = "2501.11376",
    archivePrefix = "arXiv",
    primaryClass = "hep-ph",
    month = "1",
    year = "2025"
}

@article{MoEDAL:2014ttp,
    author = "Acharya, B. and others",
    collaboration = "MoEDAL",
    title = "{The Physics Programme Of The MoEDAL Experiment At The LHC}",
    eprint = "1405.7662",
    archivePrefix = "arXiv",
    primaryClass = "hep-ph",
    reportNumber = "KCL-PH-TH-2014-02, LCTS-2014-02, CERN-PH-TH-2014-021, IFIC-14-16, IMPERIAL-TP-2014-AR-1, KCL-PH-TH/2014-02, LCTS/2014-02, CERN-PH-TH/2014-021, IFIC/14-16,
  Imperial/TP/2014/AR/1",
    doi = "10.1142/S0217751X14300506",
    journal = "Int. J. Mod. Phys. A",
    volume = "29",
    pages = "1430050",
    year = "2014"
}

@article{MoEDAL:2021vix,
    author = "Acharya, B. and others",
    collaboration = "MoEDAL",
    title = "{Search for magnetic monopoles produced via the Schwinger mechanism}",
    eprint = "2106.11933",
    archivePrefix = "arXiv",
    primaryClass = "hep-ex",
    doi = "10.1038/s41586-021-04298-1",
    journal = "Nature",
    volume = "602",
    number = "7895",
    pages = "63--67",
    year = "2022"
}

@article{Murayama:2009nj,
    author = "Murayama, Hitoshi and Shu, Jing",
    title = "{Topological Dark Matter}",
    eprint = "0905.1720",
    archivePrefix = "arXiv",
    primaryClass = "hep-ph",
    reportNumber = "UCB-PTH-09-14, IPMU09-0058",
    doi = "10.1016/j.physletb.2010.02.037",
    journal = "Phys. Lett. B",
    volume = "686",
    pages = "162--165",
    year = "2010"
}

@article{Durrer:2013pga,
    author = "Durrer, Ruth and Neronov, Andrii",
    title = "{Cosmological Magnetic Fields: Their Generation, Evolution and Observation}",
    eprint = "1303.7121",
    archivePrefix = "arXiv",
    primaryClass = "astro-ph.CO",
    doi = "10.1007/s00159-013-0062-7",
    journal = "Astron. Astrophys. Rev.",
    volume = "21",
    pages = "62",
    year = "2013"
}

@article{Graham:2017pmn,
    author = "Graham, Peter W. and Hogan, Jason M. and Kasevich, Mark A. and Rajendran, Surjeet and Romani, Roger W.",
    collaboration = "MAGIS",
    title = "{Mid-band gravitational wave detection with precision atomic sensors}",
    eprint = "1711.02225",
    archivePrefix = "arXiv",
    primaryClass = "astro-ph.IM",
    month = "11",
    year = "2017"
}

@article{Graham:2012sy,
    author = "Graham, Peter W. and Hogan, Jason M. and Kasevich, Mark A. and Rajendran, Surjeet",
    title = "{A New Method for Gravitational Wave Detection with Atomic Sensors}",
    eprint = "1206.0818",
    archivePrefix = "arXiv",
    primaryClass = "quant-ph",
    doi = "10.1103/PhysRevLett.110.171102",
    journal = "Phys. Rev. Lett.",
    volume = "110",
    pages = "171102",
    year = "2013"
}

@article{LISACosmologyWorkingGroup:2022jok,
    author = "Auclair, Pierre and others",
    collaboration = "LISA Cosmology Working Group",
    title = "{Cosmology with the Laser Interferometer Space Antenna}",
    eprint = "2204.05434",
    archivePrefix = "arXiv",
    primaryClass = "astro-ph.CO",
    reportNumber = "LISA CosWG-22-03, FERMILAB-PUB-22-349-SCD",
    doi = "10.1007/s41114-023-00045-2",
    journal = "Living Rev. Rel.",
    volume = "26",
    number = "1",
    pages = "5",
    year = "2023"
}

@article{ET:2025xjr,
    author = "Abac, Adrian and others",
    collaboration = "ET",
    title = "{The Science of the Einstein Telescope}",
    eprint = "2503.12263",
    archivePrefix = "arXiv",
    primaryClass = "gr-qc",
    reportNumber = "ET-0036C-25",
    doi = "10.1088/1475-7516/2026/03/081",
    journal = "JCAP",
    volume = "03",
    pages = "081",
    year = "2026"
}

@article{Punturo:2010zz,
    author = "Punturo, M. and others",
    editor = "Ricci, Fulvio",
    title = "{The Einstein Telescope: A third-generation gravitational wave observatory}",
    doi = "10.1088/0264-9381/27/19/194002",
    journal = "Class. Quant. Grav.",
    volume = "27",
    pages = "194002",
    year = "2010"
}

@article{Hild:2010id,
    author = "Hild, S. and others",
    title = "{Sensitivity Studies for Third-Generation Gravitational Wave Observatories}",
    eprint = "1012.0908",
    archivePrefix = "arXiv",
    primaryClass = "gr-qc",
    doi = "10.1088/0264-9381/28/9/094013",
    journal = "Class. Quant. Grav.",
    volume = "28",
    pages = "094013",
    year = "2011"
}

@article{Badurina:2019hst,
    author = "Badurina, L. and others",
    title = "{AION: An Atom Interferometer Observatory and Network}",
    eprint = "1911.11755",
    archivePrefix = "arXiv",
    primaryClass = "astro-ph.CO",
    reportNumber = "AION-2019-001, CERN-TH-2019-199",
    doi = "10.1088/1475-7516/2020/05/011",
    journal = "JCAP",
    volume = "05",
    pages = "011",
    year = "2020"
}

@article{Perri:2025qpg,
    author = "Perri, Daniele and Doro, Michele and Kobayashi, Takeshi",
    title = "{Recasting experimental constraints on relativistic magnetic monopoles}",
    eprint = "2507.05136",
    archivePrefix = "arXiv",
    primaryClass = "hep-ph",
    doi = "10.1016/j.dark.2025.102134",
    journal = "Phys. Dark Univ.",
    volume = "50",
    pages = "102134",
    year = "2025"
}

@article{Stojkovic:2004hz,
    author = "Stojkovic, Dejan and Freese, Katherine",
    title = "{A Black hole solution to the cosmological monopole problem}",
    eprint = "hep-ph/0403248",
    archivePrefix = "arXiv",
    doi = "10.1016/j.physletb.2004.12.019",
    journal = "Phys. Lett. B",
    volume = "606",
    pages = "251--257",
    year = "2005"
}

@article{Jeannerot:2003qv,
  title = {How Generic Is Cosmic String Formation in {{SUSY GUTs}}},
  author = {Jeannerot, Rachel and Rocher, Jonathan and Sakellariadou, Mairi},
  year = {2003},
  journal = {Physical Review D: Particles and Fields},
  volume = {68},
  eprint = {hep-ph/0308134},
  pages = {103514},
  doi = {10.1103/PhysRevD.68.103514},
  archiveprefix = {arXiv}
}

@article{King:2021gmj,
  title = {Confronting {{SO}}(10) {{GUTs}} with Proton Decay and Gravitational Waves},
  author = {King, Stephen F. and Pascoli, Silvia and Turner, Jessica and Zhou, Ye-Ling},
  year = {2021},
  month = oct,
  journal = {Journal of High Energy Physics},
  volume = {10},
  number = {10},
  eprint = {2106.15634},
  primaryclass = {hep-ph},
  pages = {225},
  issn = {1029-8479},
  doi = {10.1007/JHEP10(2021)225},
  urldate = {2025-04-20},
  archiveprefix = {arXiv},
  langid = {english}
}

@article{Lazarides:2023iim,
  title = {Composite {{Topological Structures}} in {{SO}}(10)},
  author = {Lazarides, George and Shafi, Qaisar and Tiwari, Amit},
  year = {2023},
  month = may,
  journal = {Journal of High Energy Physics},
  volume = {05},
  number = {5},
  eprint = {2303.15159},
  primaryclass = {hep-ph},
  pages = {119},
  issn = {1029-8479},
  doi = {10.1007/JHEP05(2023)119},
  urldate = {2025-04-06},
  archiveprefix = {arXiv},
  langid = {english}
}

@article{Parker:1970xv,
  title = {The {{Origin}} of {{Magnetic Fields}}},
  author = {Parker, E. N.},
  year = {1970},
  journal = {The Astrophysical Journal},
  volume = {160},
  pages = {383},
  issn = {0004-637X},
  doi = {10.1086/150442},
  urldate = {2025-09-20},
  langid = {english}
}

@article{Barbieri:2023qpf,
    author = "Barbieri, Riccardo and Isidori, Gino",
    title = "{Minimal flavour deconstruction}",
    eprint = "2312.14004",
    archivePrefix = "arXiv",
    primaryClass = "hep-ph",
    doi = "10.1007/JHEP05(2024)033",
    journal = "JHEP",
    volume = "05",
    pages = "033",
    year = "2024"
}

@article{Craig:2011yk,
    author = "Craig, Nathaniel and Green, Daniel and Katz, Andrey",
    title = "{(De)Constructing a Natural and Flavorful Supersymmetric Standard Model}",
    eprint = "1103.3708",
    archivePrefix = "arXiv",
    primaryClass = "hep-ph",
    reportNumber = "UMD-PP-11-002, RUNHETC-2011-06",
    doi = "10.1007/JHEP07(2011)045",
    journal = "JHEP",
    volume = "07",
    pages = "045",
    year = "2011"
}

@inproceedings{Isidori:2025iyu,
    author = "Isidori, Gino",
    title = "{Flavour Physics and CP Violation}",
    eprint = "2503.14042",
    archivePrefix = "arXiv",
    primaryClass = "hep-ph",
    month = "3",
    year = "2025"
}

@article{Isidori:2025rci,
    author = "Isidori, Gino and Paradisi, Paride and Sainaghi, Andrea and Selimovic, Nudzeim",
    title = "{Anarchic neutrinos from flavor deconstruction: phenomenology of the lepton sector}",
    eprint = "2510.23703",
    archivePrefix = "arXiv",
    primaryClass = "hep-ph",
    doi = "10.1007/JHEP02(2026)146",
    journal = "JHEP",
    volume = "02",
    pages = "146",
    year = "2026"
}

@article{Fuentes-Martin:2020bnh,
    author = "Fuentes-Mart{\'\i}n, Javier and Stangl, Peter",
    title = "{Third-family quark-lepton unification with a fundamental composite Higgs}",
    eprint = "2004.11376",
    archivePrefix = "arXiv",
    primaryClass = "hep-ph",
    reportNumber = "ZU-TH-08/20, LAPTH-014/20",
    doi = "10.1016/j.physletb.2020.135953",
    journal = "Phys. Lett. B",
    volume = "811",
    pages = "135953",
    year = "2020"
}

@article{FernandezNavarro:2022gst,
    author = "Fern{\'a}ndez Navarro, Mario and King, Stephen F.",
    title = "{B-anomalies in a twin Pati-Salam theory of flavour including the 2022 LHCb $ {R}_{K^{\left(\ast \right)}} $ analysis}",
    eprint = "2209.00276",
    archivePrefix = "arXiv",
    primaryClass = "hep-ph",
    doi = "10.1007/JHEP02(2023)188",
    journal = "JHEP",
    volume = "02",
    pages = "188",
    year = "2023"
}

@article{Capdevila:2024gki,
    author = "Capdevila, Bernat and Crivellin, Andreas and Lizana, Javier M. and Pokorski, Stefan",
    title = "{SU(2)$_{L}$ deconstruction and flavour (non)-universality}",
    eprint = "2401.00848",
    archivePrefix = "arXiv",
    primaryClass = "hep-ph",
    reportNumber = "ZU-TH 02/24, IFT-UAM/CSIC-23-162, IFT-UAM/CSIC-23-85",
    doi = "10.1007/JHEP08(2024)031",
    journal = "JHEP",
    volume = "08",
    pages = "031",
    year = "2024"
}

@article{Fuentes-Martin:2024fpx,
    author = "Fuentes-Mart{\'\i}n, Javier and Lizana, Javier M.",
    title = "{Deconstructing flavor anomalously}",
    eprint = "2402.09507",
    archivePrefix = "arXiv",
    primaryClass = "hep-ph",
    reportNumber = "IFT-UAM/CSIC-24-21",
    doi = "10.1007/JHEP07(2024)117",
    journal = "JHEP",
    volume = "07",
    pages = "117",
    year = "2024"
}

@article{FernandezNavarro:2024hnv,
    author = "Fern{\'a}ndez Navarro, Mario and King, Stephen F. and Vicente, Avelino",
    title = "{Minimal complete tri-hypercharge theories of flavour}",
    eprint = "2404.12442",
    archivePrefix = "arXiv",
    primaryClass = "hep-ph",
    doi = "10.1007/JHEP07(2024)147",
    journal = "JHEP",
    volume = "07",
    pages = "147",
    year = "2024"
}

@article{Allwicher:2023aql,
    author = "Allwicher, Lukas and Isidori, Gino and Lizana, Javier M. and Selimovic, Nudzeim and Stefanek, Ben A.",
    title = "{Third-family quark-lepton Unification and electroweak precision tests}",
    eprint = "2302.11584",
    archivePrefix = "arXiv",
    primaryClass = "hep-ph",
    reportNumber = "ZU-TH 11/23",
    doi = "10.1007/JHEP05(2023)179",
    journal = "JHEP",
    volume = "05",
    pages = "179",
    year = "2023"
}

@article{Aebischer:2022oqe,
    author = "Aebischer, Jason and Isidori, Gino and Pesut, Marko and Stefanek, Ben A. and Wilsch, Felix",
    title = "{Confronting the vector leptoquark hypothesis with new low- and high-energy data}",
    eprint = "2210.13422",
    archivePrefix = "arXiv",
    primaryClass = "hep-ph",
    reportNumber = "ZU-TH 47/22",
    doi = "10.1140/epjc/s10052-023-11304-5",
    journal = "Eur. Phys. J. C",
    volume = "83",
    number = "2",
    pages = "153",
    year = "2023"
}

@article{ATLAS:2023vxj,
    author = "Aad, Georges and others",
    collaboration = "ATLAS",
    title = "{Search for leptoquarks decaying into the b{\ensuremath{\tau}} final state in $pp$ collisions at $ \sqrt{\textrm{s}} $ = 13 TeV with the ATLAS detector}",
    eprint = "2305.15962",
    archivePrefix = "arXiv",
    primaryClass = "hep-ex",
    reportNumber = "CERN-EP-2023-033",
    doi = "10.1007/JHEP10(2023)001",
    journal = "JHEP",
    volume = "10",
    pages = "001",
    year = "2023"
}

@article{CMS:2023qdw,
    author = "Hayrapetyan, Aram and others",
    collaboration = "CMS",
    title = "{Search for a third-generation leptoquark coupled to a {\ensuremath{\tau}} lepton and a b quark through single, pair, and nonresonant production in proton-proton collisions at $ \sqrt{s} $ = 13 TeV}",
    eprint = "2308.07826",
    archivePrefix = "arXiv",
    primaryClass = "hep-ex",
    reportNumber = "CMS-EXO-19-016, CERN-EP-2023-144",
    doi = "10.1007/JHEP05(2024)311",
    journal = "JHEP",
    volume = "05",
    pages = "311",
    year = "2024"
}

@article{IceCube:2021eye,
  title = {Search for {{Relativistic Magnetic Monopoles}} with {{Eight Years}} of {{IceCube Data}}},
  author = {Abbasi, R. and Ackermann, M. and Adams, J. and Aguilar, J. A. and Ahlers, M. and Ahrens, M. and Alispach, C. and Alves, A. A. and Amin, N. M. and An, R. and Andeen, K. and Anderson, T. and Anton, G. and Arg{\"u}elles, C. and Ashida, Y. and Axani, S. and Bai, X. and Balagopal V., A. and Barbano, A. and Barwick, S. W. and Bastian, B. and Basu, V. and Baur, S. and Bay, R. and Beatty, J. J. and Becker, K.-H. and Becker Tjus, J. and Bellenghi, C. and BenZvi, S. and Berley, D. and Bernardini, E. and Besson, D. Z. and Binder, G. and Bindig, D. and Blaufuss, E. and Blot, S. and Boddenberg, M. and Bontempo, F. and Borowka, J. and B{\"o}ser, S. and Botner, O. and B{\"o}ttcher, J. and Bourbeau, E. and Bradascio, F. and Braun, J. and Bron, S. and {Brostean-Kaiser}, J. and Browne, S. and Burgman, A. and Burley, R. T. and Busse, R. S. and Campana, M. A. and {Carnie-Bronca}, E. G. and Chen, C. and Chen, Z. and Chirkin, D. and Choi, K. and Clark, B. A. and Clark, K. and Classen, L. and Coleman, A. and Collin, G. H. and Conrad, J. M. and Coppin, P. and Correa, P. and Cowen, D. F. and Cross, R. and Dappen, C. and Dave, P. and De Clercq, C. and DeLaunay, J. J. and Dembinski, H. and Deoskar, K. and Desai, A. and Desiati, P. and De Vries, K. D. and De Wasseige, G. and De With, M. and DeYoung, T. and Dharani, S. and Diaz, A. and {D{\'i}az-V{\'e}lez}, J. C. and Dittmer, M. and Dujmovic, H. and Dunkman, M. and DuVernois, M. A. and Dvorak, E. and Ehrhardt, T. and Eller, P. and Engel, R. and Erpenbeck, H. and Evans, J. and Evenson, P. A. and Fan, K. L. and Fazely, A. R. and Feigl, N. and Fiedlschuster, S. and Fienberg, A. T. and Filimonov, K. and Finley, C. and Fischer, L. and Fox, D. and Franckowiak, A. and Friedman, E. and Fritz, A. and F{\"u}rst, P. and Gaisser, T. K. and Gallagher, J. and Ganster, E. and Garcia, A. and Garrappa, S. and Gerhardt, L. and Ghadimi, A. and Glaser, C. and Glauch, T. and Gl{\"u}senkamp, T. and Gonzalez, J. G. and Goswami, S. and Grant, D. and Gr{\'e}goire, T. and Griswold, S. and G{\"u}nd{\"u}z, M. and G{\"u}nther, C. and Haack, C. and Hallgren, A. and Halliday, R. and Halve, L. and Halzen, F. and Ha Minh, M. and Hanson, K. and Hardin, J. and Harnisch, A. A. and Haungs, A. and Hauser, S. and Hebecker, D. and Helbing, K. and Henningsen, F. and Hettinger, E. C. and Hickford, S. and Hignight, J. and Hill, C. and Hill, G. C. and Hoffman, K. D. and Hoffmann, R. and Hoinka, T. and {Hokanson-Fasig}, B. and Hoshina, K. and Huang, F. and Huber, M. and Huber, T. and Hultqvist, K. and H{\"u}nnefeld, M. and Hussain, R. and In, S. and Iovine, N. and Ishihara, A. and Jansson, M. and Japaridze, G. S. and Jeong, M. and Jones, B. J. P. and Kang, D. and Kang, W. and Kang, X. and Kappes, A. and Kappesser, D. and Karg, T. and Karl, M. and Karle, A. and Katz, U. and Kauer, M. and Kellermann, M. and Kelley, J. L. and Kheirandish, A. and Kin, K. and Kintscher, T. and Kiryluk, J. and Klein, S. R. and Koirala, R. and Kolanoski, H. and Kontrimas, T. and K{\"o}pke, L. and Kopper, C. and Kopper, S. and Koskinen, D. J. and Koundal, P. and Kovacevich, M. and Kowalski, M. and Kozynets, T. and Kun, E. and Kurahashi, N. and Lad, N. and Lagunas Gualda, C. and Lanfranchi, J. L. and Larson, M. J. and Lauber, F. and Lazar, J. P. and Lee, J. W. and Leonard, K. and Leszczy{\'n}ska, A. and Li, Y. and Lincetto, M. and Liu, Q. R. and Liubarska, M. and Lohfink, E. and Lozano Mariscal, C. J. and Lu, L. and Lucarelli, F. and Ludwig, A. and Luszczak, W. and Lyu, Y. and Ma, W. Y. and Madsen, J. and Mahn, K. B. M. and Makino, Y. and Mancina, S. and Mari{\c s}, I. C. and Maruyama, R. and Mase, K. and McElroy, T. and McNally, F. and Mead, J. V. and Meagher, K. and Mechbal, S. and Medina, A. and Meier, M. and {Meighen-Berger}, S. and Micallef, J. and Mockler, D. and Montaruli, T. and Moore, R. W. and Morse, R. and Moulai, M. and Naab, R. and Nagai, R. and Naumann, U. and Necker, J. and Nguyễn, L. V. and Niederhausen, H. and Nisa, M. U. and Nowicki, S. C. and Obertacke Pollmann, A. and Oehler, M. and Oeyen, B. and Olivas, A. and O'Sullivan, E. and Pandya, H. and Pankova, D. V. and Park, N. and Parker, G. K. and Paudel, E. N. and Paul, L. and P{\'e}rez De Los Heros, C. and Peters, L. and Peterson, J. and Philippen, S. and Pieloth, D. and Pieper, S. and Pittermann, M. and Pizzuto, A. and Plum, M. and Popovych, Y. and Porcelli, A. and Prado Rodriguez, M. and Price, P. B. and Pries, B. and Przybylski, G. T. and Raab, C. and Raissi, A. and Rameez, M. and Rawlins, K. and Rea, I. C. and Rehman, A. and Reichherzer, P. and Reimann, R. and Renzi, G. and Resconi, E. and Reusch, S. and Rhode, W. and Richman, M. and Riedel, B. and Roberts, E. J. and Robertson, S. and Roellinghoff, G. and Rongen, M. and Rott, C. and Ruhe, T. and Ryckbosch, D. and Rysewyk Cantu, D. and Safa, I. and Saffer, J. and Sanchez Herrera, S. E. and Sandrock, A. and Sandroos, J. and Santander, M. and Sarkar, S. and Sarkar, S. and Satalecka, K. and Scharf, M. and Schaufel, M. and Schieler, H. and Schindler, S. and Schlunder, P. and Schmidt, T. and Schneider, A. and Schneider, J. and Schr{\"o}der, F. G. and Schumacher, L. and Schwefer, G. and Sclafani, S. and Seckel, D. and Seunarine, S. and Sharma, A. and Shefali, S. and Silva, M. and Skrzypek, B. and Smithers, B. and Snihur, R. and Soedingrekso, J. and Soldin, D. and Spannfellner, C. and Spiczak, G. M. and Spiering, C. and Stachurska, J. and Stamatikos, M. and Stanev, T. and Stein, R. and Stettner, J. and Steuer, A. and Stezelberger, T. and St{\"u}rwald, T. and Stuttard, T. and Sullivan, G. W. and Taboada, I. and Tenholt, F. and {Ter-Antonyan}, S. and Tilav, S. and Tischbein, F. and Tollefson, K. and Tomankova, L. and T{\"o}nnis, C. and Toscano, S. and Tosi, D. and Trettin, A. and Tselengidou, M. and Tung, C. F. and Turcati, A. and Turcotte, R. and Turley, C. F. and Twagirayezu, J. P. and Ty, B. and Unland Elorrieta, M. A. and {Valtonen-Mattila}, N. and Vandenbroucke, J. and Van Eijndhoven, N. and Vannerom, D. and Van Santen, J. and Verpoest, S. and Walck, C. and Watson, T. B. and Weaver, C. and Weigel, P. and Weindl, A. and Weiss, M. J. and Weldert, J. and Wendt, C. and Werthebach, J. and Weyrauch, M. and Whitehorn, N. and Wiebusch, C. H. and Williams, D. R. and Wolf, M. and Woschnagg, K. and Wrede, G. and Wulff, J. and Xu, X. W. and Yanez, J. P. and Yoshida, S. and Yu, S. and Yuan, T. and Zhang, Z. and {IceCube Collaboration}},
  year = {2022},
  month = feb,
  journal = {Physical Review Letters},
  volume = {128},
  number = {5},
  eprint = {2109.13719},
  primaryclass = {astro-ph.HE},
  pages = {051101},
  issn = {0031-9007, 1079-7114},
  doi = {10.1103/PhysRevLett.128.051101},
  urldate = {2025-09-20},
  archiveprefix = {arXiv},
  collaboration = {IceCube},
  langid = {english}
}

@article{ParticleDataGroup:2024cfk,
  title = {Review of Particle Physics},
  author = {Navas, S. and others},
  year = {2024},
  month = aug,
  journal = {Phys. Rev. D},
  volume = {110},
  number = {3},
  pages = {030001},
  doi = {10.1103/PhysRevD.110.030001},
  collaboration = {Particle Data Group}
}

@article{Bordone:2017bld,
  title = {A Three-Site Gauge Model for Flavor Hierarchies and Flavor Anomalies},
  author = {Bordone, Marzia and Cornella, Claudia and {Fuentes-Mart{\'i}n}, Javier and Isidori, Gino},
  year = {2018},
  month = apr,
  journal = {Physics Letters B},
  volume = {779},
  eprint = {1712.01368},
  primaryclass = {hep-ph},
  pages = {317--323},
  issn = {03702693},
  doi = {10.1016/j.physletb.2018.02.011},
  urldate = {2025-09-20},
  archiveprefix = {arXiv},
  langid = {english}
}

@article{Schwinger:1951nm,
    author = "Schwinger, Julian S.",
    editor = "Milton, K. A.",
    title = "{On gauge invariance and vacuum polarization}",
    doi = "10.1103/PhysRev.82.664",
    journal = "Phys. Rev.",
    volume = "82",
    pages = "664--679",
    year = "1951"
}

@article{Ho:2021uem,
    author = "Ho, David L. -J. and Rajantie, Arttu",
    title = "{Instanton solution for Schwinger production of {\textquoteright}t Hooft-Polyakov monopoles}",
    eprint = "2103.12799",
    archivePrefix = "arXiv",
    primaryClass = "hep-th",
    reportNumber = "IMPERIAL-TP-2021-DH-04",
    doi = "10.1103/PhysRevD.103.115033",
    journal = "Phys. Rev. D",
    volume = "103",
    number = "11",
    pages = "115033",
    year = "2021"
}

@article{Affleck:1981ag,
    author = "Affleck, Ian K. and Manton, Nicholas S.",
    title = "{Monopole Pair Production in a Magnetic Field}",
    reportNumber = "MIT-CTP-933",
    doi = "10.1016/0550-3213(82)90511-9",
    journal = "Nucl. Phys. B",
    volume = "194",
    pages = "38--64",
    year = "1982"
}

@article{PierreAuger:2016imq,
  title = {Search for Ultrarelativistic Magnetic Monopoles with the {{Pierre Auger}} Observatory},
  author = {Aab, Alexander and Abreu, Pedro and Aglietta, Marco and others},
  year = {2016},
  month = oct,
  journal = {Physical Review D},
  volume = {94},
  number = {8},
  eprint = {1609.04451},
  primaryclass = {astro-ph.HE},
  pages = {082002},
  publisher = {American Physical Society},
  doi = {10.1103/PhysRevD.94.082002},
  urldate = {2025-09-20},
  archiveprefix = {arXiv},
  collaboration = {Pierre Auger}
}

@article{Froggatt:1978nt,
    author = "Froggatt, C. D. and Nielsen, Holger Bech",
    title = "{Hierarchy of Quark Masses, Cabibbo Angles and CP Violation}",
    reportNumber = "CERN-TH-2519",
    doi = "10.1016/0550-3213(79)90316-X",
    journal = "Nucl. Phys. B",
    volume = "147",
    pages = "277--298",
    year = "1979"
}

@article{Arkani-Hamed:2001nha,
    author = "Arkani-Hamed, Nima and Cohen, Andrew G. and Georgi, Howard",
    title = "{Electroweak symmetry breaking from dimensional deconstruction}",
    eprint = "hep-ph/0105239",
    archivePrefix = "arXiv",
    reportNumber = "HUTP-01-A024, BUHEP-01-06, UCB-PTH-01-15",
    doi = "10.1016/S0370-2693(01)00741-9",
    journal = "Phys. Lett. B",
    volume = "513",
    pages = "232--240",
    year = "2001"
}

@article{DAmbrosio:2002vsn,
    author = "D'Ambrosio, G. and Giudice, G. F. and Isidori, G. and Strumia, A.",
    title = "{Minimal flavor violation: An Effective field theory approach}",
    eprint = "hep-ph/0207036",
    archivePrefix = "arXiv",
    reportNumber = "CERN-TH-2002-147, IFUP-TH-2002-17",
    doi = "10.1016/S0550-3213(02)00836-2",
    journal = "Nucl. Phys. B",
    volume = "645",
    pages = "155--187",
    year = "2002"
}

@article{Allwicher:2023shc,
    author = "Allwicher, Lukas and Cornella, Claudia and Isidori, Gino and Stefanek, Ben A.",
    title = "{New physics in the third generation. A comprehensive SMEFT analysis and future prospects}",
    eprint = "2311.00020",
    archivePrefix = "arXiv",
    primaryClass = "hep-ph",
    reportNumber = "ZU-TH 71/23, MITP-23-060, KCL-PH-TH/2023-59",
    doi = "10.1007/JHEP03(2024)049",
    journal = "JHEP",
    volume = "03",
    pages = "049",
    year = "2024"
}

@article{Covone:2024elw,
    author = "Covone, Sebastiano and Davighi, Joe and Isidori, Gino and Pesut, Marko",
    title = "{Flavour deconstructing the composite Higgs}",
    eprint = "2407.10950",
    archivePrefix = "arXiv",
    primaryClass = "hep-ph",
    reportNumber = "CERN-TH-2024-112",
    doi = "10.1007/JHEP01(2025)041",
    journal = "JHEP",
    volume = "01",
    pages = "041",
    year = "2025"
}

@article{Fuentes-Martin:2022xnb,
    author = "Fuentes-Martin, Javier and Isidori, Gino and Lizana, Javier M. and Selimovic, Nudzeim and Stefanek, Ben A.",
    title = "{Flavor hierarchies, flavor anomalies, and Higgs mass from a warped extra dimension}",
    eprint = "2203.01952",
    archivePrefix = "arXiv",
    primaryClass = "hep-ph",
    reportNumber = "ZU-TH-08/22",
    doi = "10.1016/j.physletb.2022.137382",
    journal = "Phys. Lett. B",
    volume = "834",
    pages = "137382",
    year = "2022"
}

@article{Barbieri:2021wrc,
    author = "Barbieri, Riccardo",
    title = "{A View of Flavour Physics in 2021}",
    eprint = "2103.15635",
    archivePrefix = "arXiv",
    primaryClass = "hep-ph",
    doi = "10.5506/APhysPolB.52.789",
    journal = "Acta Phys. Polon. B",
    volume = "52",
    number = "6-7",
    pages = "789",
    year = "2021"
}

@article{Allwicher:2020esa,
    author = "Allwicher, Lukas and Isidori, Gino and Thomsen, Anders Eller",
    title = "{Stability of the Higgs Sector in a Flavor-Inspired Multi-Scale Model}",
    eprint = "2011.01946",
    archivePrefix = "arXiv",
    primaryClass = "hep-ph",
    reportNumber = "ZU-TH-41/20",
    doi = "10.1007/JHEP01(2021)191",
    journal = "JHEP",
    volume = "01",
    pages = "191",
    year = "2021"
}

@article{Panico:2016ull,
    author = "Panico, Giuliano and Pomarol, Alex",
    title = "{Flavor hierarchies from dynamical scales}",
    eprint = "1603.06609",
    archivePrefix = "arXiv",
    primaryClass = "hep-ph",
    reportNumber = "CERN-TH-2016-065",
    doi = "10.1007/JHEP07(2016)097",
    journal = "JHEP",
    volume = "07",
    pages = "097",
    year = "2016"
}

@article{Dvali:2000ha,
    author = "Dvali, G. R. and Shifman, Mikhail A.",
    title = "{Families as neighbors in extra dimension}",
    eprint = "hep-ph/0001072",
    archivePrefix = "arXiv",
    reportNumber = "TPI-MINN-00-03-T, UMN-TH-1836-00, NYU-TH-00-01-01",
    doi = "10.1016/S0370-2693(00)00083-6",
    journal = "Phys. Lett. B",
    volume = "475",
    pages = "295--302",
    year = "2000"
}

@article{Barbieri:2012uh,
    author = "Barbieri, Riccardo and Buttazzo, Dario and Sala, Filippo and Straub, David M.",
    title = "{Flavour physics from an approximate $U(2)^3$ symmetry}",
    eprint = "1203.4218",
    archivePrefix = "arXiv",
    primaryClass = "hep-ph",
    doi = "10.1007/JHEP07(2012)181",
    journal = "JHEP",
    volume = "07",
    pages = "181",
    year = "2012"
}

@article{Allwicher:2024ncl,
    author = "Allwicher, Lukas and Bordone, Marzia and Isidori, Gino and Piazza, Gioacchino and Stanzione, Alfredo",
    title = "{Probing third-generation New Physics with K{\textrightarrow}{\ensuremath{\pi}}{\ensuremath{\nu}}{\ensuremath{\nu}}{\textasciimacron} and B{\textrightarrow}K({\textasteriskcentered}){\ensuremath{\nu}}{\ensuremath{\nu}}{\textasciimacron}}",
    eprint = "2410.21444",
    archivePrefix = "arXiv",
    primaryClass = "hep-ph",
    reportNumber = "CERN-TH-2024-183",
    doi = "10.1016/j.physletb.2025.139295",
    journal = "Phys. Lett. B",
    volume = "861",
    pages = "139295",
    year = "2025"
}

@article{Greljo:2018tuh,
    author = "Greljo, Admir and Stefanek, Ben A.",
    title = "{Third family quark{\textendash}lepton unification at the TeV scale}",
    eprint = "1802.04274",
    archivePrefix = "arXiv",
    primaryClass = "hep-ph",
    reportNumber = "MITP-18-012",
    doi = "10.1016/j.physletb.2018.05.033",
    journal = "Phys. Lett. B",
    volume = "782",
    pages = "131--138",
    year = "2018"
}

@article{Li:1981nk,
    author = "Li, Xiaoyuan and Ma, Ernest",
    title = "{Gauge Model of Generation Nonuniversality}",
    reportNumber = "Print-81-0777 (HAWAII), UH-511-454-81",
    doi = "10.1103/PhysRevLett.47.1788",
    journal = "Phys. Rev. Lett.",
    volume = "47",
    pages = "1788",
    year = "1981"
}

@article{Barbieri:2011ci,
    author = "Barbieri, Riccardo and Isidori, Gino and Jones-Perez, Joel and Lodone, Paolo and Straub, David M.",
    title = "{$U(2)$ and Minimal Flavour Violation in Supersymmetry}",
    eprint = "1105.2296",
    archivePrefix = "arXiv",
    primaryClass = "hep-ph",
    doi = "10.1140/epjc/s10052-011-1725-z",
    journal = "Eur. Phys. J. C",
    volume = "71",
    pages = "1725",
    year = "2011"
}

@article{Fuentes-Martin:2019mun,
    author = "Fuentes-Mart{\'\i}n, Javier and Isidori, Gino and Pag{\`e}s, Julie and Yamamoto, Kei",
    title = "{With or without U(2)? Probing non-standard flavor and helicity structures in semileptonic B decays}",
    eprint = "1909.02519",
    archivePrefix = "arXiv",
    primaryClass = "hep-ph",
    reportNumber = "ZU-TH-42/19",
    doi = "10.1016/j.physletb.2019.135080",
    journal = "Phys. Lett. B",
    volume = "800",
    pages = "135080",
    year = "2020"
}

@article{Redi:2012uj,
    author = "Redi, Michele",
    title = "{Composite MFV and Beyond}",
    eprint = "1203.4220",
    archivePrefix = "arXiv",
    primaryClass = "hep-ph",
    reportNumber = "CERN-PH-TH-2012-067",
    doi = "10.1140/epjc/s10052-012-2030-1",
    journal = "Eur. Phys. J. C",
    volume = "72",
    pages = "2030",
    year = "2012"
}

@article{Isidori:2012ts,
    author = "Isidori, Gino and Straub, David M.",
    title = "{Minimal Flavour Violation and Beyond}",
    eprint = "1202.0464",
    archivePrefix = "arXiv",
    primaryClass = "hep-ph",
    doi = "10.1140/epjc/s10052-012-2103-1",
    journal = "Eur. Phys. J. C",
    volume = "72",
    pages = "2103",
    year = "2012"
}

@article{Allwicher:2025bub,
    author = "Allwicher, Lukas and Isidori, Gino and Pesut, Marko",
    title = "{Flavored circular collider: cornering New Physics at FCC-ee via flavor-changing processes}",
    eprint = "2503.17019",
    archivePrefix = "arXiv",
    primaryClass = "hep-ph",
    reportNumber = "DESY-25-046",
    doi = "10.1140/epjc/s10052-025-14359-8",
    journal = "Eur. Phys. J. C",
    volume = "85",
    number = "6",
    pages = "631",
    year = "2025"
}

@book{Naber:1997yu,
  title = {Topology, {{Geometry}} and {{Gauge}} Fields: {{Foundations}}},
  shorttitle = {Topology, {{Geometry}} and {{Gauge}} Fields},
  author = {Naber, Gregory L.},
  year = 1997,
  series = {Texts {{Appl}}. {{Math}}.},
  volume = {25},
  publisher = {Springer New York},
  address = {New York, NY},
  doi = {10.1007/978-1-4419-7254-5},
  urldate = {2025-04-16},
  copyright = {https://www.springernature.com/gp/researchers/text-and-data-mining},
  isbn = {978-1-4419-7253-8 978-1-4419-7254-5},
  langid = {english}
}

@article{Guth:1980zm,
  title = {Inflationary Universe: {{A}} Possible Solution to the Horizon and Flatness Problems},
  shorttitle = {Inflationary Universe},
  author = {Guth, Alan H.},
  year = 1981,
  journal = {Physical Review D},
  volume = {23},
  number = {2},
  pages = {347--356},
  issn = {0556-2821},
  doi = {10.1103/PhysRevD.23.347},
  urldate = {2024-11-10},
  copyright = {http://link.aps.org/licenses/aps-default-license},
  langid = {english}
}

@article{Linde:1981mu,
  title = {A New Inflationary Universe Scenario: {{A}} Possible Solution of the Horizon, Flatness, Homogeneity, Isotropy and Primordial Monopole Problems},
  shorttitle = {A New Inflationary Universe Scenario},
  author = {Linde, A. D.},
  year = 1982,
  journal = {Physics Letters B},
  volume = {108},
  number = {6},
  pages = {389--393},
  issn = {0370-2693},
  doi = {10.1016/0370-2693(82)91219-9},
  urldate = {2025-11-05}
}

@article{Kibble:1976sj,
  title = {Topology of Cosmic Domains and Strings},
  author = {Kibble, T W B},
  year = 1976,
  journal = {Journal of Physics A: Mathematical and General},
  volume = {9},
  number = {8},
  pages = {1387--1398},
  issn = {0305-4470, 1361-6447},
  doi = {10.1088/0305-4470/9/8/029},
  urldate = {2025-02-16},
  langid = {english}
}

@article{Polyakov:1974ek,
    author = "Polyakov, Alexander M.",
    editor = "Taylor, J. C.",
    title = "{Particle Spectrum in Quantum Field Theory}",
    reportNumber = "PRINT-74-1566 (LANDAU-INST)",
    journal = "JETP Lett.",
    volume = "20",
    pages = "194--195",
    year = "1974"
}

@article{Bogomolny:1976ab,
  title = {Calculation of the Monopole Mass in Gauge Theory},
  author = {Bogomolny, E. B. and Marinov, M. S.},
  year = 1976,
  journal = {Yadernaya Fizika},
  volume = {23},
  pages = {676--680}
}

@article{Zurek:1985qw,
  title = {Cosmological Experiments in Superfluid Helium?},
  author = {Zurek, W. H.},
  year = 1985,
  journal = {Nature},
  volume = {317},
  number = {6037},
  pages = {505--508},
  publisher = {Nature Publishing Group},
  issn = {1476-4687},
  doi = {10.1038/317505a0},
  urldate = {2025-05-04},
  copyright = {1985 Springer Nature Limited},
  langid = {english}
}

@article{Preskill:1984gd,
    author = "Preskill, John",
    title = "{MAGNETIC MONOPOLES}",
    reportNumber = "CALT-68-1108",
    doi = "10.1146/annurev.ns.34.120184.002333",
    journal = "Ann. Rev. Nucl. Part. Sci.",
    volume = "34",
    pages = "461--530",
    year = "1984"
}

@article{Perri:2023ncd,
  title = {Monopole Acceleration in Intergalactic Magnetic Fields},
  author = {Perri, Daniele and Bondarenko, Kyrilo and Doro, Michele and Kobayashi, Takeshi},
  year = 2024,
  month = oct,
  journal = {Physics of the Dark Universe},
  volume = {46},
  eprint = {2401.00560},
  primaryclass = {hep-ph},
  pages = {101704},
  issn = {22126864},
  doi = {10.1016/j.dark.2024.101704},
  url = {https://linkinghub.elsevier.com/retrieve/pii/S2212686424002863},
  urldate = {2025-10-19},
  archiveprefix = {arXiv},
  langid = {english}
}

@article{Zeldovich:1978wj,
  title = {On the Concentration of Relic Magnetic Monopoles in the Universe},
  author = {Zeldovich, {\relax Ya.B}. and Khlopov, {\relax M.Yu}.},
  year = 1978,
  journal = {Physics Letters B},
  volume = {79},
  number = {3},
  pages = {239--241},
  issn = {03702693},
  doi = {10.1016/0370-2693(78)90232-0},
  url = {https://linkinghub.elsevier.com/retrieve/pii/0370269378902320},
  urldate = {2025-09-23},
  copyright = {https://www.elsevier.com/tdm/userlicense/1.0/},
  langid = {english}
}

@article{Turner:1982ag,
  title = {Magnetic Monopoles and the Survival of Galactic Magnetic Fields},
  author = {Turner, Michael S. and Parker, E. N. and Bogdan, T. J.},
  year = 1982,
  journal = {Physical Review D},
  volume = {26},
  number = {6},
  pages = {1296},
  issn = {0556-2821},
  doi = {10.1103/PhysRevD.26.1296},
  url = {https://link.aps.org/doi/10.1103/PhysRevD.26.1296},
  urldate = {2025-09-22},
  copyright = {http://link.aps.org/licenses/aps-default-license},
  langid = {english}
}

@book{Mimura:2000xxx,
  title = {Topology of {{Lie Groups}}, {{I}} and {{II}}},
  author = {Mimura, Mamoru and Toda, Hirosi},
  year = 2000,
  month = aug,
  series = {Translations of {{Mathematical Monographs}}},
  volume = {91},
  publisher = {American Mathematical Society},
  address = {Providence, Rhode Island},
  doi = {10.1090/mmono/091},
  url = {https://www.ams.org/mmono/091},
  urldate = {2026-05-18},
  isbn = {978-0-8218-1342-3 978-0-8218-8761-5 978-1-4704-4503-4},
  langid = {english}
}
\end{document}